\pdfoutput=1
\documentclass[11pt,twoside,a4paper,cmspaper,final,collab]{cms-tdr}
\def\svnVersion{8db8f0f}\def\svnDate{2026/06/16}\def\cmsCernNoTag{CERN-EP-2025-119}\def\cmsCernDate{\today}\def\cmsMessage{Published in the Journal of High Energy Physics as \href{http://dx.doi.org/10.1007/JHEP06(2026)014}{\doi{10.1007/JHEP06(2026)014}.}}
\begin{document}\cmsNoteHeader{SUS-23-018}

\cmsNoteHeader{SUS-23-018}

\newlength\cmsFigWidth
\setlength\cmsFigWidth{0.49\textwidth}
\newlength\cmsTabSkip\setlength{\cmsTabSkip}{2ex}

\newcommand{\pp}{\ensuremath{\Pp\Pp}\xspace}
\newcommand{\mH}{\ensuremath{m_\PH}\xspace}
\newcommand{\mh}{\ensuremath{m_\Ph}\xspace}
\newcommand{\mHpm}{\ensuremath{m_\PHpm}\xspace}
\newcommand{\mA}{\ensuremath{m_{\PA}}\xspace}
\newcommand{\ma}{\ensuremath{m_\Pa}\xspace}
\newcommand{\mchi}{\ensuremath{m_\PGc}\xspace}
\newcommand{\mZ}{\ensuremath{m_\PZ}\xspace}
\newcommand{\BR}{\ensuremath{\mathcal{B}}\xspace}
\providecommand{\PGgst}{\ensuremath{\HepParticle{\PGg}{}{\ast}}\xspace}
\providecommand{\PAGc}{\ensuremath{\HepAntiParticle{\PGc}{}{}}\xspace}
\providecommand{\PA}{\ensuremath{\HepParticle{A}{}{}}\xspace}
\renewcommand{\Pl}{\ensuremath{l}\xspace}
\renewcommand{\PAl}{\ensuremath{\bar{l}\xspace}}

\title{Search for dark matter production in association with bottom quarks and a lepton pair in proton-proton collisions at \texorpdfstring{$\sqrt{s}=13\TeV$}{sqrt(s)=13 TeV}}

\date{\today}

\abstract{A search is performed for dark matter produced in association with bottom quarks and a pair of electrons or muons in data collected with the CMS detector at the LHC, corresponding to 138\fbinv of integrated luminosity of proton-proton collisions at a center-of-mass energy of 13\TeV. For the first time at the LHC, the associated production of a bottom quark-antiquark pair and a new heavy neutral Higgs boson (\PH) that subsequently decays into a leptonically decaying \PZ boson and a pseudoscalar (\Pa) is explored. The latter acts as a dark matter mediator in the context of the two Higgs doublet model plus a pseudoscalar (2HDM+\Pa). Multivariate techniques that target a wide range of mass configurations for the \PH and \Pa particles are used. The observations are consistent with the expectations from standard model processes. Upper limits at 95\% confidence level are set on the product of cross section and branching fraction of the new particles, ranging from $10^{-2}$\unit{pb} for an \PH mass of 400\GeV to $10^{-3}$\unit{pb} for an \PH mass of 2000\GeV. Constraints on the parameter space of a benchmark 2HDM+\Pa model are derived and compared with expectations in the context of cosmological predictions.}

\hypersetup{
pdfauthor={CMS Collaboration},
pdftitle={Search for dark matter production in association with bottom quarks and a lepton pair in proton-proton collisions at sqrt(s)=13 TeV},
pdfsubject={CMS},
pdfkeywords={CMS, dark matter, heavy scalar, galactic center excess, leptons, b-quarks}} 

\maketitle 

\section{Introduction}
\label{sec:intro}

The nature of dark matter (DM) is unknown. A prevalent hypothesis suggests that the DM could exist in the form of a weakly interacting massive particle (WIMP)~\cite{Steigman:1984ac,Roszkowski_2018}. Many WIMP candidates, hereafter referred to simply as ``DM particles'', are expected to have been thermally produced in the early universe, similar to the particles of the standard model (SM). If the mass of the DM particle is in the \GeV-\TeVns range, the respective relic abundance can be obtained via thermal freeze-out in the early universe~\cite{Tunney:2017yfp,Bertone:2004pz}. 

An extensive experimental program is currently ongoing to search for DM particles through interactions with SM particles. Three main approaches broadly cover the efforts of DM detection: direct searches, indirect searches, and searches at colliders. Direct searches aim at detecting the scattering of ambient DM particles from nuclei, while in indirect searches the energetic particles possibly produced by DM annihilations in space are analyzed. Complementary to these two approaches, searches for the production of DM particles at particle colliders created under well-controlled conditions are also of high relevance. Since DM particles would escape detection, their presence would lead to events involving an imbalance in the vector sum of the measured transverse momenta through the presence of missing transverse momentum (\ptmiss) recoiling against a visible final state $\PX$. Searches for events with large \ptmiss are currently a major focus at CERN's LHC~\cite{CMS:2024zqs,DeRoeck:2024fjq,ATLAS:2024kpy,ATLAS:2023rvb}.

Recently, the Fermi-LAT space telescope observed a gamma-ray excess in studies of the Milky Way galactic center~\cite{Fermi-LAT:2015sau}. 
While the consistency with other observations in astroparticle physics and cosmology is still under discussion~\cite{MAGIC:2016xys}, this might be interpreted as the existence of weak-scale DM annihilating into bottom quark-antiquark (\bbbar) pairs~\cite{Hooper:2010mq,Hooper:2011ti,Abazajian:2012pn,Karwin:2016tsw}.
A compelling interpretation of the gamma-ray excess from the galactic center, in agreement with the fact that no DM particle has been observed in direct detection experiments so far, is given by the existence of a pseudoscalar mediator between the SM and DM sectors~\cite{Boehm:2014hva,Izaguirre:2014vva,Ipek:2014gua}. This would lead to spin-dependent DM-nucleon interactions, for which experimental limits are much less stringent than for spin-independent interactions~\cite{Berlin:2014tja,Arina:2014yna,XENON:2019rxp,dycz8zf}. In general, the strength of DM-nucleon interactions mediated by pseudoscalars is below the reach of present direct DM detection experiments~\cite{Boveia:2016mrp,Berlin:2015wwa}.

In this paper, we present a novel DM search for events involving final states with a \bbbar pair, a leptonically decaying \PZ boson, and \ptmiss. The search is performed using data recorded with the CMS detector at the LHC in proton-proton (\pp) collisions at a center-of-mass energy of 13\TeV. Such a signature could be produced in scenarios beyond the SM (BSM) with a pseudoscalar providing the link between visible SM particles and the dark sector. This signature is particularly interesting as it might be sensitive to unexplored regions of the parameter space where searches for DM~\cite{ATLAS:2021kxv,CMS:2021far,ATLAS:2020uiq,CMS:2018ffd,ATLAS:2021gcn,CMS:2020ulv,ATLAS:2021shl,ATLAS:2021jbf,CMS:2019ykj,CMS:2018zjv,CMS:2018nlv,ATLAS:2022bzt,ATLAS:2020fgc,ATLAS:2022znu,ATLAS:2020yzc,CMS:2018ysw,CMS:2019zzl} and flavor bounds~\cite{CMS:2022mgd,ATLAS:2018cur,Misiak:2017bgg,Belle:2016ufb,Logan:2000iv,Misiak:2015xwa,CMS:2013dcn} can not reach. Additionally, it can serve as a test of the DM interpretation of the gamma-ray galactic center excess~\cite{Tunney:2017yfp}.

One alternative to generate a gauge-invariant and renormalizable model featuring a pseudoscalar DM mediator is to introduce an additional Higgs doublet, as typically done in the so-called two Higgs doublet models (2HDMs)~\cite{Nomura:2008ru,Goncalves:2016iyg,No:2015xqa,Branco:2011iw}. 
In this scenario, the coupling of a DM mediator to SM fermions is naturally generated by introducing a new pseudoscalar field that mixes with the 2HDM pseudoscalar field. This new field couples directly to the DM candidate, thus naturally providing a connection between the SM and DM sectors.
The resulting model is referred to as 2HDM+$\Pa$~\cite{Bauer:2017ota}, and has been thoroughly studied in the context of the LHC Dark Matter Working Group~\cite{LHCDarkMatterWorkingGroup:2018ufk}. The scalar sector of the 2HDM+$\Pa$ contains a charged scalar \Hpm, two neutral 2HDM CP-even scalars \PH and \Ph, and two neutral 2HDM CP-odd scalars $\PA$ and $\Pa$. 
The mixing between the pseudoscalar components allows both \PA and \Pa to couple simultaneously to DM and SM fermions, providing the portal between the visible and DM sectors. The angle that controls this mixing is called $\theta$. 
The ratio of the vacuum expectation values of the two doublets is given by \tanb, while the mixing angle $\alpha$ in the neutral CP-even sector satisfies $\beta-\alpha = \pi/2$. 

\begin{figure}[htbp]
\centering
\includegraphics[width=0.3\textwidth]{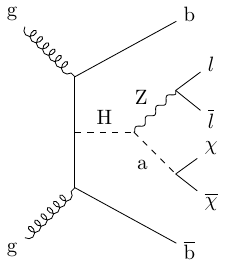}
\caption{Example diagram at leading order for the production of a heavy pseudoscalar mediator decaying into dark matter particles, in association with $\PZ(\to \Pl\PAl) \bbbar$. }
\label{Fig:Feynman_Graph}
\end{figure}

An additional incentive to further examine the 2HDM+$\Pa$ is that this theoretical construction is able to correctly predict the DM relic density when DM is annihilated via an s-channel process into SM fermions. 
When the mass of the DM particles is chosen to match the mass range preferred by the gamma-ray excess, the observed DM relic density tends to favor relatively large values of \tanb~\cite{Tunney:2017yfp}.
Larger values of this parameter lead to larger couplings of the heavy scalar states to down-type fermions, such as bottom quarks.
This configuration motivates the search channel investigated in this paper, namely heavy \PH production in association with a \bbbar pair ($\bbbar\PH$), where the heavy scalar and the new pseudoscalar particles decay as $\PH \to \PZ\Pa \to (\Pl\PAl)(\PGc \PAGc)$ in which \Pl denotes an electron (\Pe), a muon (\PGm), or a tau (\PGt) lepton and \PGc and \PAGc a DM particle and antiparticle, respectively. 
Although \Pl can also be a \PGt lepton, the analysis only focuses on final states with a pair of electrons or muons.
An example diagram for the process described above is shown in Fig.~\ref{Fig:Feynman_Graph}.
Searching for this process at the LHC allows probing the region of parameter space consistent with a DM interpretation of the gamma-ray galactic center excess, not accessible through other processes. For example, Ref.~\cite{Tunney:2017yfp} demonstrates that this channel provides feasible access to the intermediate $\tanb$ region ($\tanb \approx 10$), while simultaneously probing mixing values down to $\sin\theta \approx 0.1$.
In this article, we present the first dedicated experimental analysis of this channel in the context of searches for DM at colliders. 

This article is organized as follows. After a brief description of the CMS detector in Section~\ref{sec:cms}, the data and simulated samples are described in Section~\ref{sec:samples}. In Section~\ref{sec:reco}, the event reconstruction is outlined, while Section~\ref{sec:selection} describes the event selection, including the multivariate analysis. The background estimation is discussed in Section~\ref{sec:background}, while the systematic uncertainties are given in Section~\ref{sec:systematics}. The final results are presented in Section~\ref{sec:results} and a summary is given in Section~\ref{sec:summary}. Tabulated results are provided in the HEPData record for this analysis \cite{hepdata}.

\section{The CMS detector}
\label{sec:cms}

The central feature of the CMS apparatus is a superconducting solenoid of 6\unit{m} internal diameter, providing a magnetic field of 3.8\unit{T}. Within the solenoid volume are a silicon pixel and strip tracker, a lead tungstate crystal electromagnetic calorimeter (ECAL), and a brass and scintillator hadron calorimeter (HCAL), each composed of a barrel and two endcap sections. Forward calorimeters extend the pseudorapidity ($\eta$) coverage provided by the barrel and endcap detectors. Muons are measured in gas-ionization detectors embedded in the steel flux-return yoke outside the solenoid. More detailed descriptions of the CMS detector, together with a definition of the coordinate system used and the relevant kinematic variables, can be found in Ref.~\cite{CMS:2008xjf}.

Events of interest are selected using a two-tiered trigger system. The first level, composed of custom hardware processors, uses information from the calorimeters and muon detectors to select events at a rate of around 100\unit{kHz} within a fixed latency of 4\mus~\cite{CMS:2020cmk}. The second level, known as the high-level trigger, consists of a farm of processors running a version of the full event reconstruction software optimized for fast processing, and reduces the event rate to around 1\unit{kHz} before data storage~\cite{CMS:2016ngn}.

\section{Data and simulated samples}
\label{sec:samples}

This search uses data recorded in 2016--2018 with the CMS detector, corresponding to an integrated luminosity of 138\fbinv. 
The data sets used in the analysis have been selected by a lepton trigger, which is composed of a combined logical OR of single-lepton and dilepton triggers. 
The dilepton triggers have isolation requirements and transverse momentum ($\pt$) thresholds of 17--23 and 8--12\GeV for the leading and subleading leptons, respectively.
The inclusion of single-lepton triggers with lepton $\pt$ threshold higher than 23\GeV and no isolation requirement is important to select signal events with two energetic leptons spatially close to each other that occur more frequently as $\mH - \ma$ increases. 
This results in high trigger efficiency compared to isolated lepton triggers when the leptons are reconstructed in each other's isolation cones.
The lepton trigger efficiencies are measured in data and simulation using independent triggers based on \ptmiss. The average trigger efficiency, evaluated across lepton $\pt$ and $\eta$ bins, is 98.6\%. Correction factors are derived and applied to the simulated samples to match the trigger efficiencies measured in the data.

The signal process is simulated at leading order (LO) using \MGvATNLO (MG5) v2.6.5~\cite{Alwall:2014hca} with a dedicated 2HDM+$\Pa$ model~\cite{Bauer:2017ota}. The benchmark scenario investigated is summarized as

\begin{equation}
\begin{split}
    & \mH = \mA = \mHpm, \quad\mh = 125\GeV, \quad\mchi=45\GeV, \\
    & \cos(\beta-\alpha)=0, \quad\tanb = 11, \quad\sin\theta = 0.35, \\
    & \lambda_{3}=\lambda_{P_{1}}=\lambda_{P_{2}}=0.25, \quad\text{and} \quad y_{\PGc} = 1,\\
\end{split}
\label{Eq:BenchmarkChoice}
\end{equation}

where \mH, \mA, and \mHpm are the masses of heavy Higgs bosons, while the light neutral CP-even scalar $\Ph$ is identified with the Higgs boson. The mass of the pseudoscalar mediator and the DM particle are denoted with \ma and \mchi, respectively.  The $\lambda_{P_{1}}$ and $\lambda_{P_{2}}$ parameters control the quartic interactions between the doublets and the singlet, $\lambda_{3}$ mediates the same quartic interaction but only between the two doublets, and $y_\PGc$ is the parameter that controls the strength of the interaction between the singlet scalar and the DM fermion $\PGc$. The couplings of \PA and \Pa to DM particles are given by $y_\PGc\cos\theta $ and $y_\PGc\sin\theta$, respectively.
A thorough study of various constraints implied by the parameter choices of Eq.~\eqref{Eq:BenchmarkChoice} is presented in Ref.~\cite{LHCDarkMatterWorkingGroup:2018ufk}. 
The value of \mchi is chosen according to Refs.~\cite{Tunney:2017yfp,Karwin:2016tsw}, where it is shown that such a value lies close to the preferred range ($\mchi \in [50, 170]\GeV$) of the gamma-ray galactic-center excess if DM annihilates into \bbbar. Its exact value is irrelevant for this search when it is below half the mass of the mediator. Different signal samples are generated for the scalar masses $\mH$ and $\ma$, taken as free parameters, with $\mH (> \ma + \mZ)$ between 400 and 2000\GeV and $\ma (> 2 \mchi)$ between 100 and 1800\GeV. The $\lambda_3$ parameter is set to $\lambda_3 = m^{2}_{\Ph} /v^{2} \approx 0.25$, with $v$ the electroweak vacuum expectation value. In this configuration, the contribution from the $\PH \to \Pa\Pa$ channel becomes insignificant, as shown in Ref.~\cite{Bauer:2017ota}, thus guaranteeing a sufficiently narrow width of the heavy scalar resonance. The \PH boson width has been calculated for relevant variations of the other 2HDM+$\Pa$ parameters for this analysis, verifying the validity of the narrow-width approximation for both new resonances. 

The SM processes with the largest contributions to the background in this analysis are Drell--Yan (DY) $\PZ/\PGgst \to \Pl\PAl$, top quark-antiquark pair (\ttbar), single top quark, and diboson (in particular $\PW\PZ$ and $\PZ\PZ$) production. For the $\PZ\PZ$ case, if one of the $\PZ$ bosons decays to a charged lepton pair while the other decays to a pair of neutrinos, this results in an irreducible signature if an initial-state radiation jet is identified as \PQb jet. Similarly for the $\PW\PZ$ process, if the charged lepton from the \PW boson decay is lost or not identified, this can result in a signal-like signature.

Samples of $\PZ/\PGgst \to \Pl\PAl$ events are generated at next-to-leading order (NLO) in quantum chromodynamics (QCD) using MG5 with the FxFx prescription~\cite{Frederix:2012ps} for matching and merging of two additional jets from the matrix element (ME) calculation to the parton shower. Simulated events of \ttbar, single top quark production in the $t$ channel and in association with a \PW boson ($\PQt\PW$), $\PW\PZ$, and $\PZ\PZ$ are generated at NLO in QCD using \POWHEG~v2~\cite{Frixione:2007vw,Alioli:2010xd}. Single top quark production in the $s$ channel is simulated at NLO in QCD using MG5. Additional minor SM contributions from $\PV\PV$, $\PV\PV\PV$, $\ttbar\PV$, $\ttbar\PV\PV$, and $\PQt\PZ\Pq$ processes, in which \PV stands for \PW or \PZ, will be referred to as ``Other'' and are simulated with MG5 at NLO or LO in QCD, depending on the process.

All background samples are normalized using the most accurate cross section calculations available, which generally incorporate NLO or next-to-NLO (NNLO) precision~\cite{Li:2012wna,Czakon:2011xx,Kidonakis:2010ux,Campbell:2011bn}. 
It was also observed that the $\pt$ spectra of top quarks in $\ttbar$ data were significantly softer than those predicted by simulations based on either LO or NLO matrix elements interfaced with parton showers. The simulation can be improved considering a correction based on the latest theoretical NNLO QCD + NLO electroweak (EW) calculation for the SM $\ttbar$ production \cite{Czakon:2017wor}. The correction was implemented by a reweighting procedure with correction factors depending on the \pt of the top quark and antiquark. It is applied to the events of the NLO sample simulating the $\ttbar$ process. 
Similarly, for the $\PW\PZ$ process, a correction based on NNLO QCD + NLO EW fixed-order predictions obtained using {{\textsc{matrix}}\xspace v2.0.0.beta1}~\cite{Grazzini:2017mhc,CMS:2021icx} has been incorporated via reweighting the simulated WZ sample with correction factors depending on the \pt of the W boson. For all the above-described simulations, the initial-state partons are modeled with the NNPDF 3.1 NNLO~\cite{NNPDF:2017mvq} parton distribution functions (PDFs). Parton showering, hadronization, and the underlying event dynamics are handled by {\PYTHIA~v8.240}~\cite{Sj_strand_2015} using the \textsc{CP5} tune~\cite{CMS:2020wcf}. All signal and background samples are processed using \GEANTfour~\cite{GEANT4:2002zbu} to provide a full simulation of the CMS detector, including a simulation of the triggers.

The effects of additional \pp interactions in the same or adjacent bunch crossings, referred to as pileup, are included in all simulation samples. A reweighting procedure is used to match the simulated distribution of pileup interactions with the one observed in data.

\section{Event reconstruction}
\label{sec:reco}

The particle-flow (PF) algorithm~\cite{CMS:2017yfk} aims to reconstruct and identify each individual particle in an event, with an optimized combination of information from the various elements of the CMS detector including charged particle tracks from the tracking detector, energy deposits in the HCAL and ECAL, and reconstructed tracks from the muon chambers. Particles in each event are reconstructed and identified as either electrons, muons, photons, charged hadrons, or neutral hadrons. 
The primary vertex is taken to be the vertex corresponding to the hardest scattering in the event, selected as the vertex with the highest quadratic sum of the \pt from all tracks associated with it~\cite{Contardo:2020886}.

Electrons are identified as a primary track and potentially multiple ECAL energy clusters corresponding to the extrapolation of this track to the ECAL and to possible bremsstrahlung photons emitted along the way. The energy of electrons is deduced from the electron momentum at the PV, as determined by the tracker, the energy of the corresponding ECAL cluster, and the cumulative energy from all bremsstrahlung photons which have their origin spatially compatible with the electron track. 
An identification algorithm~\cite{CMS:2020uim} is applied using multivariate techniques and including isolation information during the training. A working point with an identification efficiency of 90\% and a misidentification rate of 1\% for jets misidentified as electrons is used. 
Only electrons with $\pt > 20\GeV$, $\abs{\eta} < 2.5$, and $\abs{\eta_\mathrm{SC}} \notin [1.444,1.566]$ are selected, where $\eta_\mathrm{SC}$ is the $\eta$ of the ECAL cluster corresponding to the electron candidate. 
The $\eta_\mathrm{SC}$ requirement aims to avoid a gap between the ECAL barrel and endcap components. Corrections for electron reconstruction and identification efficiencies are applied to simulated events to improve the agreement between data and simulation~\cite{CMS:2020uim}.

Muons are identified as tracks in the tracker consistent with either a track or several hits in the muon system, and associated with calorimeter deposits compatible with the muon hypothesis. The momentum of muons is obtained from the curvature of the corresponding track. In this analysis, muons must be isolated from charged hadrons and neutral particles, and pass the ``medium'' identification criteria~\cite{CMS:2018rym} optimized for prompt muons and muons from heavy flavor decay, having an identification efficiency of 98\% and misidentification rates of pions and kaons as muons below 0.2\% and 0.5\%, respectively. Muons are required to have $\pt>20\GeV$ and $\abs{\eta}<2.4$. Corrections for muon reconstruction, identification, and isolation efficiencies are applied to simulated events to improve the agreement with data~\cite{CMS:2018rym}.

The PF candidates are clustered into jets using the anti-\kt clustering algorithm~\cite{Cacciari:2008gp}, with a distance parameter of 0.4, implemented in the \FASTJET package~\cite{Cacciari:2011ma,Cacciari:2005hq}. Charged particles reconstructed from tracks associated with pileup vertices are omitted from the clustering to reduce the impact of pileup collisions, while charged particles associated with the primary vertex and all neutral particles are kept.
The jet energy is determined from the vectorial sum of all particle four-momenta in the jet and is found from simulation to be, on average, within 5--10\% of the true momentum over the entire \pt spectrum and detector acceptance. 
Jet energy corrections are derived from simulation studies so that the average measured energy of reconstructed jets becomes identical to that of particle-level jets. Measurements using dijet, photon+jet, \PZ{}+jet, and multijet events are used to determine any residual differences between the jet energy scale in data and simulation~\cite{CMS:2016lmd}, and appropriate corrections are made. 
The energy resolution of simulated jets is also corrected to match the resolution in the data, and typically amounts to 15--20\% at 30\GeV, 10\% at 100\GeV, and 5\% at 1\TeV~\cite{CMS:2016lmd}. 

Jets in the angular vicinity of a selected electron or muon with $\DR < 0.4$ are removed from the analysis, where $\DR = \sqrt{(\Delta\eta)^2 + (\Delta\phi)^2}$ with $\Delta\eta$ and $\Delta\phi$ as the distances in pseudorapidity and azimuth, respectively, between the lepton and the jet. 
Jets must have $\pt > 20$\GeV and $\abs{\eta} < 2.4$ to be considered in this analysis. 
Jets with $\pt < 50$\GeV are required to pass the loose working point criteria of the pileup jet identification~\cite{CMS:2020ebo} to suppress jets from pileup interactions.
The \textsc{DeepCSV} algorithm~\cite{CMS:2017wtu} is used to identify jets originating from the fragmentation of a \PQb quark (\PQb jets), at a loose working point defined by an inclusive light-quark or gluon jet misidentification rate of 10\%, resulting in a \PQb-jet identification efficiency of 84\%.
Additional corrections are applied to cover remaining residual differences between data and simulation that arise from pileup and b-tagged jet identification efficiencies.

The missing transverse momentum vector \ptvecmiss is computed for each event as the negative vector sum of the transverse momenta of the reconstructed physics objects, namely muons, electrons, photons, hadronically decaying \PGt leptons, jets, and unclustered energy~\cite{CMS:2019ctu}. The unclustered energy is the contribution from the PF candidates not associated with any of the previous physics objects. The \ptvecmiss is modified to account for corrections to the energy scale and resolution of reconstructed jets in the event. Its magnitude, denoted as \ptmiss, is an important observable in this analysis because it provides information related to the DM candidates in signal events. Collision events can exhibit unusually large values of \ptmiss which may be attributed to reconstruction failures, detector malfunctions, or noncollision backgrounds. To address these issues, filters are designed to detect and reject events with spurious \ptmiss values, having an efficiency of 85\%--90\% and a misidentification rate of less than 0.1\%~\cite{CMS:2019ctu}. These filters are applied to data and simulated events.

\section{Event selection}
\label{sec:selection}

The signal topology in this analysis consists of a pair of leptons from a \PZ boson decay with opposite electrical charge and same flavor, the presence of b jets, and a substantial amount of \ptmiss from the $\Pa \to \chi \bar{\chi}$ decay. The event selection starts with a sequence of requirements in kinematic variables referred to as baseline selection, which are applied in all regions defined in the analysis. Additional requirements are imposed to separate the remaining events into the signal region (SR) and control regions (CRs).

\subsection{Baseline selection}
\label{Subsec:Baseline}

The selection requires at least one jet and two opposite-sign (OS) leptons ($\Pell_1$ and $\Pell_2$) where only $\Pem\Pep$, $\PGmm\PGmp$, $\Pem\PGmp$, and $\PGmm\Pep$ pairs are considered. 
If there are more than two leptons in an event, the chosen OS dilepton system is the one with the reconstructed invariant mass closer to \mZ, assumed to be 91.19\GeV~\cite{Workman:2022ynf}. The lepton with the highest \pt must have $\pt^{\Pell_1} > 40\GeV$.
The dilepton system must have transverse momentum $\pt^{\Pell\Pell} > 40\GeV$, an angular distance between the two leptons $\DR^{\Pell\Pell} < 3.2$, and a small mass difference 
$\Delta m^{\Pell\Pell} = \abs{ m_{\Pell\Pell} - \mZ } < 25\GeV$, intended to reduce the number of events from nonresonant sources. 
The requirement $\ptmiss > 65\GeV$ removes SM backgrounds without high-\pt neutrinos in the final state. 
In signal events, the \PZ boson and the pseudoscalar particle are more likely to be emitted back-to-back in $\phi$. 
The difference in azimuthal angle between the dilepton system and \ptvecmiss must satisfy $\Delta\phi^{\Pell\Pell,\ptmiss} > 0.8$.
A quantity that is related to the mass of the heavy scalar is the transverse mass $\mT^{\Pell\Pell,\ptmiss}$ of the system formed by the dilepton system and \ptvecmiss~\cite{LESTER199999}. 
Events are required to have $\mT^{\Pell\Pell,\ptmiss}>90\GeV$.

The baseline selection removes more than 97\% of the events from each of the main background processes. 
The impact on the signals is a reduction in selection efficiency by about 10\% from the $N_\text{jets} \geq 1$ requirement, and 3--8\% from the other requirements depending on the signal sample.

\subsection{Signal region selection}
\label{Subsec:SR}

Because the two \PQb quarks produced in signal events typically have low \pt, many of the corresponding jets do not pass the $\pt > 20$\GeV criterion applied at event reconstruction. 
To maximize the acceptance, only a loose \PQb-jet requirement is adopted for the SR that consists of the presence of at least one \PQb jet with the loose \PQb tagging working point ($N_\PQb \geq 1$) defined in Section \ref{sec:reco}.
Additionally, the events must have exactly two opposite-sign same-flavor (OSSF) leptons ($\Pep\Pem$ or $\PGmp\PGmm$).

To suppress the contamination of \ttbar production in events with two reconstructed leptons and \PQb jets, an analytical method is used to reconstruct and identify such events~\cite{PhysRevD.73.054015}. This method makes use of an algebraic approach to solve a system of equations with six kinematic constraints to determine the unknown momentum vectors of the two undetected neutrinos. The constraint equations are constructed using the four momenta of the two leptons, two jets, and \ptvecmiss, assuming the top quark and \PW masses to be 172.5 and 80.37\GeV, respectively~\cite{Workman:2022ynf}. 
Events with an analytic solution to this set of equations are assumed to be from \ttbar production. 

The \ttbar event reconstruction is performed on events with at least two jets and one b-tagged jet.  If there are more than two jets in an event, the algorithm tries to find a solution as follows.
\begin{itemize}
\item In events with two or more b-tagged jets, all pairs of b-tagged jets are used in the reconstruction. If solutions exist, these are considered by the algorithm.
\item If no solution is found in the previous hypothesis or if the event has only one b-tagged jet, all pairs of a b-tagged jet with another jet are used in the reconstruction. Again, if solutions exist, these are considered by the algorithm.
\end{itemize}
For any of the above cases, the generator level distributions of the neutrino energy ($E_{\nu}$) and antineutrino energy ($E_{\bar{\nu}}$), as obtained from simulated \ttbar production, are used as probability distributions. The best solution is the one with the reconstructed values of $E_{\nu}$ and $E_{\bar{\nu}}$ associated with the highest probability. In case there is no solution, the event is considered to have not originated from \ttbar production.

The non-existence of an analytical solution for the \ttbar event reconstruction described above is added as the last requirement for the SR selection. By doing that, about 60\% of the \ttbar dilepton events are removed from the SR, while reducing the signal efficiency by 0.2--5\%, depending on the signal sample. A list of all requirements used to define the SR is given in Table~\ref{Tab:baselineAndSR}.

\begin{table}[htbp]
\centering
\topcaption{List of requirements used to define the SR, split into a baseline and an SR region selection.}
\label{Tab:baselineAndSR}
\begin{tabular}{ll}
\hline
Baseline selection & SR selection \\
\hline
Has two OS leptons & Pass baseline selection \\
$\pt^{\Pell_1} >$ 40\GeV & $N_{\PQb} \ge 1$ \\
$\ptmiss >$ 65\GeV & $N_\mathrm{leptons} = 2$ \\
$\pt^{\Pell\Pell} >$ 40\GeV & $\Pell_1$ and $\Pell_2$ are OSSF \\
$\Delta m^{\Pell\Pell} <$ 25\GeV & No \ttbar analytic solution \\
$\DR^{\Pell\Pell} <$ 3.2 & \\
$\Delta\phi^{\Pell\Pell,\ptmiss} >$ 0.8 & \\
$\mT^{\mathrm{\Pell\Pell},\ptmiss} >$ 90\GeV & \\
$N_\text{jets} \geq$ 1 &  \\
\hline
\end{tabular}
\end{table}

The majority of $\PZ\PZ$ events in the SR come from $\PZ\PZ \to 2\Pell2\PGn$, while for $\PW\PZ$ events, the main channel is $\PW\PZ \to 3\Pell\PGn$ with one lepton not reconstructed. Except for $\PZ\PZ \to 2\Pell2\Pq$, the $\PW\PZ$, $\PZ\PZ$, and DY events present in the SR are typically there because of misidentification of light-flavor, c quark, and gluon jets as b jets, since their matrix element does not have \PQb quarks in its final state. In general, for each one of these processes, more than 80\% of the events in the SR are there due to the misidentification of other jets as \PQb jets. Naturally, this fraction is larger for the loose \PQb tagging working point compared to stricter \PQb tagging requirements.

\subsection{Multivariate optimization}
\label{Subsec:MVA}

Fully connected multi-layer artificial neural networks, also known as multi-layer perceptrons (MLPs), are used to combine variables based on \ptvecmiss and information from the leptons into a single discriminant to sort the events in the SR into intervals of low to high sensitivity. The MLPs are implemented using the PyTorch package~\cite{NEURIPS2019_9015} and are trained on simulated events in the SR for each data-taking period. 

The following variables are passed to the MLPs as input: $\pt^{\Pell_1}$, $\pt^{\Pell_2}$, $\pt^{\Pell\Pell}$, $\DR^{\Pell\Pell}$, $\Delta m^{\Pell\Pell}$, \ptmiss, $\mT^{\mathrm{\Pell\Pell},\ptmiss}$, $\Delta\phi^{\Pell\Pell,\ptmiss}$, and $m_\mathrm{T2}^{\Pell\Pell}$.
The latter variable is useful to discriminate events from the decay of a primary particle into an invisible and a visible particle~\cite{LESTER199999,Lester2015}. 
It is related to the mass difference between the primary and the invisible particles,
\begin{equation}
m_\mathrm{T2}^{\Pell\Pell}=\underset{\vec{p}_{\mathrm{T},\PAGn}+\vec{p}_{\mathrm{T},\PGn}=\ptvecmiss} 
{\min}\left[ \max\{ \mT (m_{\Pell},m_{\PAGn},\vec{p}_{\mathrm{T},\Pell},\vec{p}_{\mathrm{T},\PAGn} ), 
\mT (m_{\overline{\Pell}},m_{\PGn},\vec{p}_{\mathrm{T},\overline{\Pell}},\vec{p}_{\mathrm{T},\PGn})\} \right],
\label{Eq:Sec5:MT2ll}
\end{equation}
with
\begin{equation}
\mT (m_{1},m_{2},\vec{p}_{\mathrm{T},1},\vec{p}_{\mathrm{T},2}) 
= \sqrt{ m^2_{1} + m^2_{2} + 2 (E_{\mathrm{T},1} E_{\mathrm{T},2} - \vec{p}_{\mathrm{T},1} \cdot \vec{p}_{\mathrm{T},2})},
\label{Eq:Sec5:MT}
\end{equation}
where $m_{i}$, $\vec{p}_{\mathrm{T},i}$, and $E_{\mathrm{T},i}$ are the mass, transverse momentum vector, and transverse energy of particle $i$, respectively.

Variables directly related to jet kinematics were not used as input to the MLP discriminator to avoid propagating to the MLP score their shape mismodeling. The incorrect modeling was observed in the CRs when comparing the MC distributions of these variables with those from data. Jet mismodeling may occur due to the intrinsic complexity in reconstructing a jet and due to neutrinos coming from hadronic decays inside the jet.
The inclusion of \ptmiss information in the input variables was important since it has a large discrimination power between signal and background events.
However, since the \ptmiss scale and resolution in DY simulation are more affected by jet mismodeling than in other processes, they needed to be corrected to match those in the data, as described in Section~\ref{sec:background}. 

During the training, the contributions from all processes are normalized to the integrated luminosity of the data and expected cross sections.
The signal class groups events from all signal samples, while the background class groups events from all background processes.  
For each class, the sum of weights is then normalized to the same value to achieve equal statistical importance for both classes during the MLP training.

\begin{figure}[htbp!]
\centering
\includegraphics[width=0.49\textwidth]{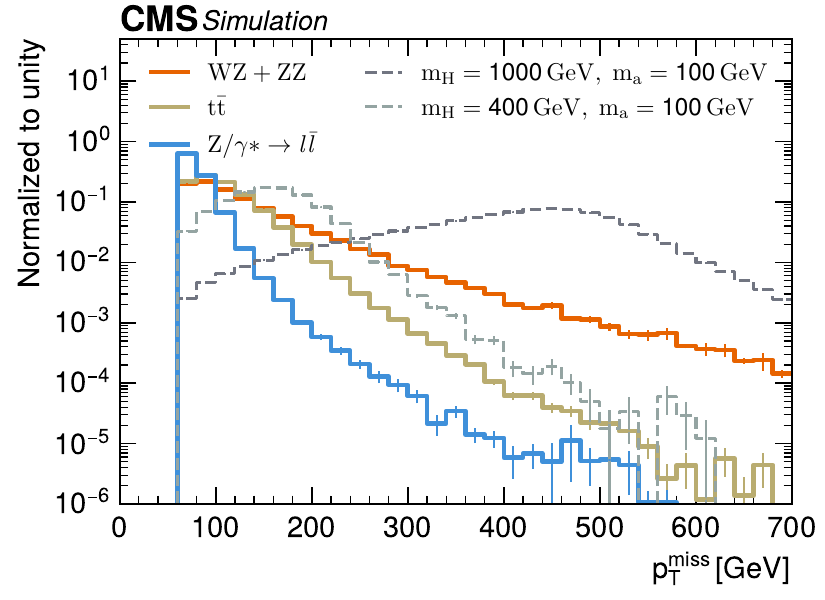}
\includegraphics[width=0.49\textwidth]{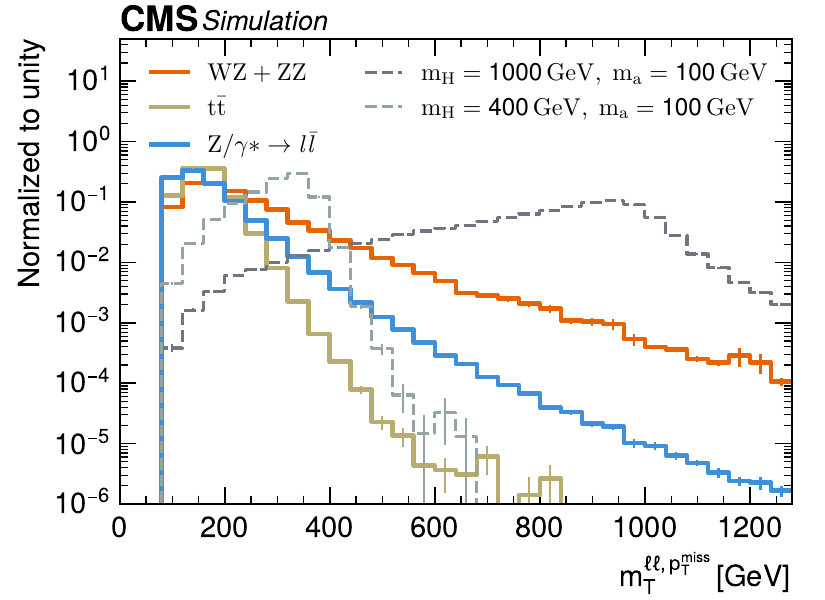}
\includegraphics[width=0.49\textwidth]{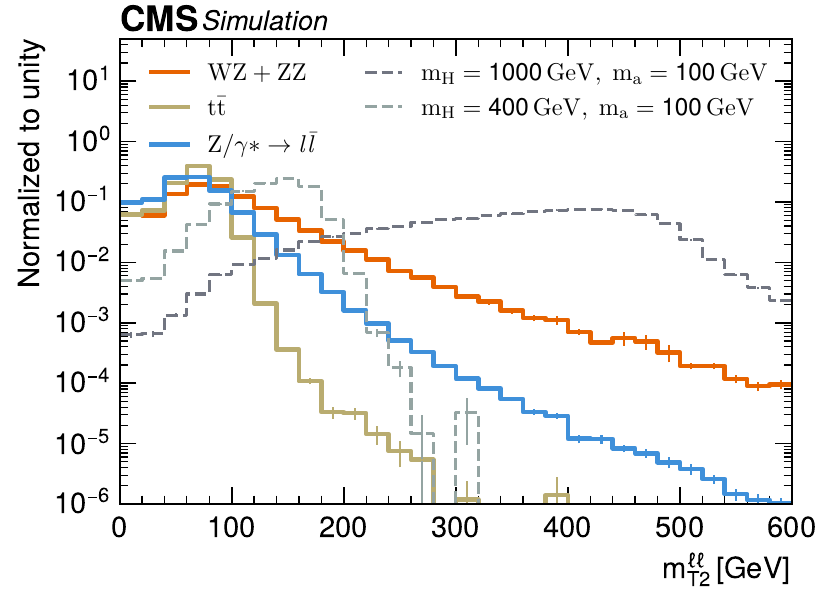}
\includegraphics[width=0.49\textwidth]{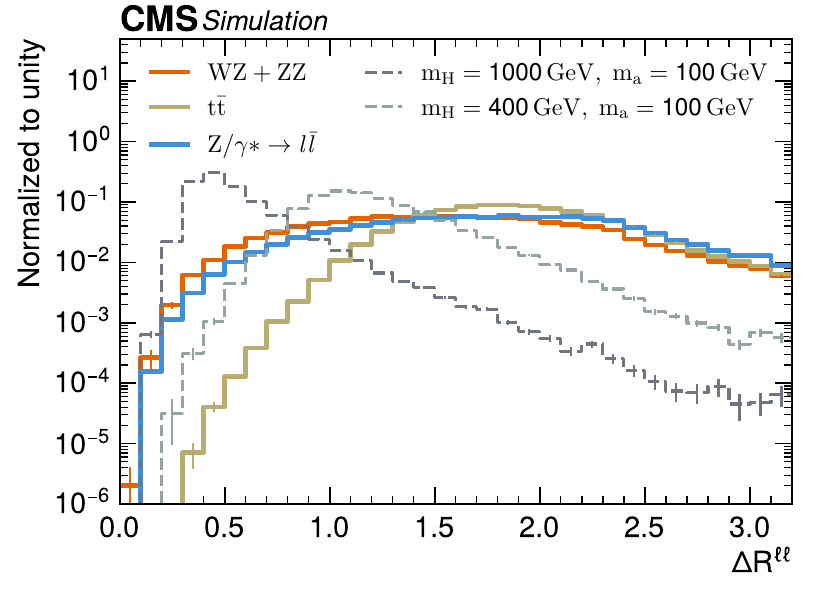}
\caption{Normalized distributions in \ptmiss (upper left), $\mT^{\mathrm{\Pell\Pell},\ptmiss}$ (upper right), $m_\mathrm{T2}^{\Pell\Pell}$ (lower left), and $\DR^{\Pell\Pell}$ (lower right) in the SR for the main background processes (solid lines) and two signal samples representing extreme kinematic behaviors, which are obtained by considering high (dark gray dashed line) and low (light gray dashed line) values of \mH and $\mH-\ma$. The vertical bars at the center of the bins represent the statistical uncertainty in the predictions.}
\label{Fig:Best_variables}
\end{figure}

The MLP architectures with the best discrimination performance for each data-taking period are composed of two or three hidden layers and 20 nodes in each hidden layer. The input variables that contribute most to the discrimination performance are found to be \ptmiss and $\mT^{\mathrm{\Pell\Pell},\ptmiss}$, followed by $m_\mathrm{T2}^{\Pell\Pell}$ and $\DR^{\Pell\Pell}$, estimated using information from the permutation importance method \cite{Breiman2001}. Figure~\ref{Fig:Best_variables} shows a comparison between signal and background distributions in these variables in the SR, normalized to unit area. Since the $\PW\PZ$ contribution comes from events with one lepton not reconstructed, its kinematic distributions are similar to the ones from $\PZ\PZ$.
For this reason, these two backgrounds are combined in Fig.~\ref{Fig:Best_variables}.

To construct the search intervals used to extract the signal, the MLP score is transformed into the MLP4 score, defined as
\begin{equation}
\mathrm{MLP4\ score} = \frac{10^{4\times\mathrm{MLP\ score}}}{10^4}.
\end{equation}
The MLP4 score, primarily based on statistical considerations, expands the region with a high sensitivity making the intervals in MLP4 more practical and effective. 

In addition, we observe that the MLP4 score returns values close to 1 for more boosted signal samples (higher $\pt^{\Pell\Pell}$), while less boosted signal samples (lower $\pt^{\Pell\Pell}$) show a peak at slightly smaller values in the MLP4 score distribution. 
To maximize the sensitivity for both types of signal events, while preserving the smoothness of the distributions, we define the search intervals in the MLP4 score as
\begin{equation*}
[0.00,0.01,0.08,0.15,0.22,0.29,0.36,0.43,0.50,0.57,0.64,0.71,0.78,0.85,0.90,0.95,0.98,1.00],
\end{equation*}
in which smaller intervals are defined in the region closest to 1 in order to be sensitive to the shape variation of the different signal samples. Below 0.85, equal-size intervals are defined, except for the first interval, which is equal to 0.01 to avoid a large concentration of events there, most of them being background.
The MLP4 bins will be labeled with integer numbers ranging from 1 to 17, indicating the position of the corresponding interval in the above definition.

\section{Background estimation}
\label{sec:background}

The analysis relies partially on simulation to estimate the background contribution from SM processes. 
The $\ttbar$ and $\PW\PZ$ simulations are improved by applying NNLO(+NLO) corrections, as described in Section~\ref{sec:samples}. 
In turn, the \ptmiss scale and resolution in the DY simulation are corrected using an auxiliary measurement derived in a sideband region with $\ptmiss < 65\GeV$ and extrapolated to the main analysis region with larger values of \ptmiss.
The measurement consists of propagating to \ptmiss the correction of the parallel and orthogonal hadronic recoil projections on the direction of the reconstructed dilepton transverse momentum vector ($\vec{\pt}^{\Pell\Pell}$). The correction is derived for different jet multiplicities and $\pt^{\Pell\Pell}$ intervals using a parameterized double Gaussian function to match the simulation distributions to the ones in data. 
The extrapolation is verified in a region with no \PQb jets and $\ptmiss < 300\GeV$, also dominated by DY events. 

\begin{figure}[htbp]
\includegraphics[width=1\textwidth]{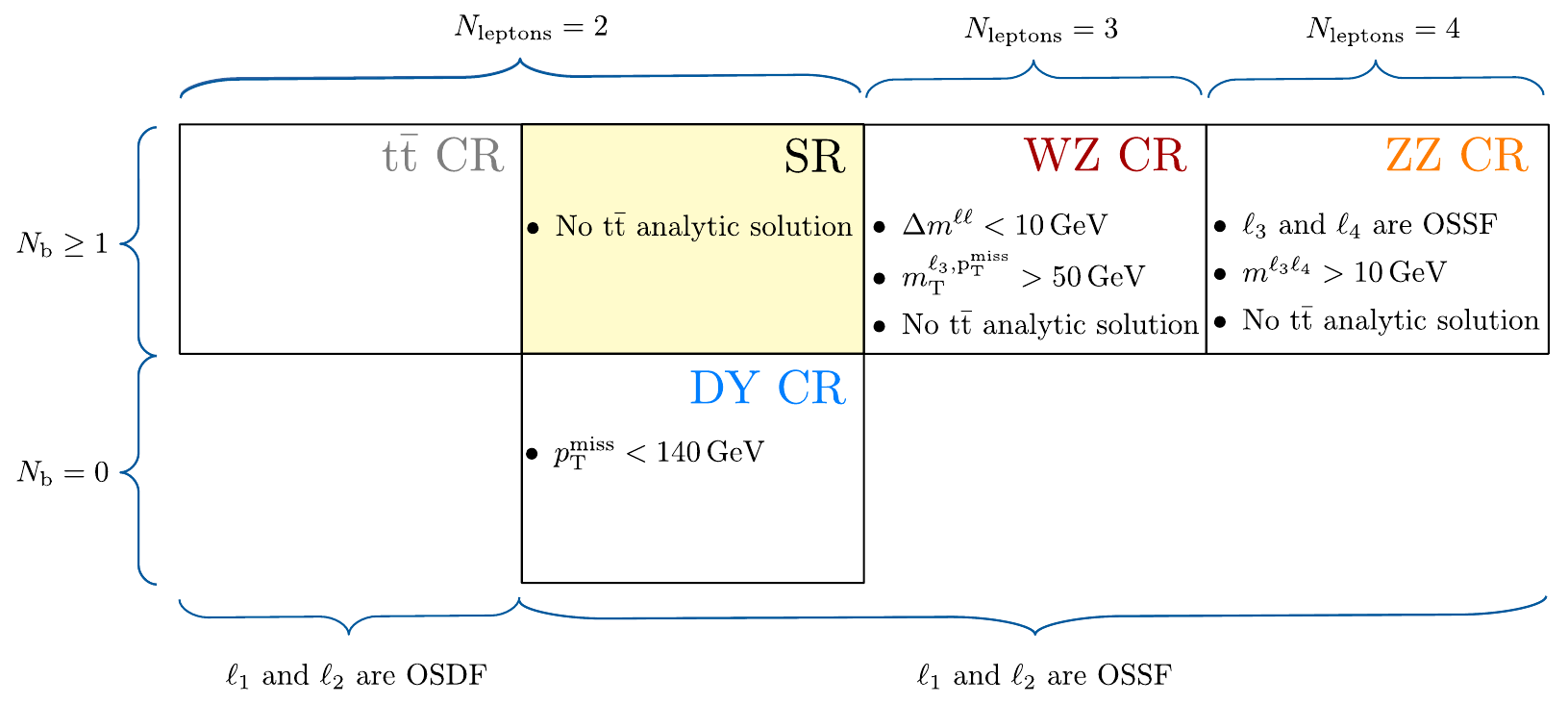}
\caption{Illustration of the requirements on the SR and CRs. 
The regions are represented as blocks, and the requirements are denoted 
by curly brackets placed around the blocks and bullets within each block.
All requirements are applied on top of the baseline selection.}
\label{Fig:regions}
\end{figure}

The normalization of the largest backgrounds in this search, 
namely DY, \ttbar, $\PW\PZ$, and $\PZ\PZ$ production, is corrected using a data-driven method. It makes use of unconstrained parameters associated with the normalization of each background process that are linked across various regions in a maximum likelihood fit to data (discussed in Section~\ref{sec:results}).
The definitions of the CRs are described below. Their corresponding purities are 95\% (DY CR), 92\% (\ttbar CR), 62\% ($\PW\PZ$ CR), and 95\% ($\PZ\PZ$ CR). We show a summary of the SR and CRs in Fig.~\ref{Fig:regions}, 
which illustrates the variables used to define the different regions.

All CRs, except for the DY CR, must have $N_{\PQb} \ge 1$, identical to the SR, where $N_{\PQb}$ denotes the number of \PQb jets. These regions contain a significant fraction of \ttbar events. 
To reduce the contributions from \ttbar production in the $\PW\PZ$ and $\PZ\PZ$ CRs, we discard events that have an analytic solution for the dileptonic \ttbar event reconstruction described in Section~\ref{Subsec:SR}. Additional requirements defining the CRs are listed below.

\begin{itemize}
\item As in the SR, events in the DY CR must have exactly two OSSF leptons. Events with $\ptmiss > 140\GeV$ are discarded to suppress a potential contamination from signal and other background processes. Events in the DY CR must have $N_{\PQb}=0$. We checked the consistency between the normalization of DY events with $N_{\PQb}=0$ and $N_{\PQb} \geq 1$ by comparing the ratio of the number of data and simulation events in additional regions with lower values of \ptmiss ($< 40\GeV$ and between 40--65$\GeV$). We observe ratios that are compatible between the $N_{\PQb}=0$ and $N_{\PQb} \geq 1$ categories, validating the use of the DY CR for correcting the DY normalization in the SR.
\item In the \ttbar CR, events must have exactly two opposite-sign different-flavor (OSDF) leptons, \ie $\Pep\PGmm$ or $\Pem\PGmp$, resulting in a region enriched with \ttbar events in the dilepton channel. 
\item In the $\PW\PZ$ CR, events must have exactly two OSSF leptons and one additional lepton $\Pell_3$. To improve the purity of this CR in $\PW\PZ$ events, we add the following requirements in well-reconstructed variables, $\Delta m^{\Pell\Pell} < 10\GeV$ and $\mT^{\mathrm{\Pell_3},\ptmiss} > 50\GeV$, where $\mT^{\mathrm{\Pell_3},\ptmiss}$ is the transverse mass of the $\Pell_3$ and \ptvecmiss system. 
\item In the $\PZ\PZ$ CR, besides the two OSSF leptons with invariant mass closer to \mZ, two additional OSSF leptons, $\Pell_3$ and $\Pell_4$, are required. In addition, a condition on the reconstructed invariant mass of this extra dilepton system is imposed to remove contamination from other processes with negligible impact on the amount of $\PZ\PZ$ events, namely $m^{\Pell_3 \Pell_4} > 10\GeV$. 
\end{itemize}

Intrinsically, the DY process does not have undetectable particles with large \pt in the final state, such that events with large \ptmiss are rare. Nonnegligible values of \ptmiss can originate from detector acceptance effects, as well as mismeasurements. 
For this reason, DY events contribute mostly to the region of small \ptmiss and constitute an important background for signal samples with small $\pt^{\Pell\Pell}$. 
On the other hand, $\PZ\PZ$ and $\PW\PZ$ production with large \ptmiss are important backgrounds for signal samples with large $\pt^{\Pell\Pell}$. 
In the SR, 91\% of the $\PZ\PZ$ events are those in which the second $\PZ$ boson does not decay to charged leptons, while 66\% of the $\PW\PZ$ events have a $\PW$ boson decaying to one neutrino and one charged lepton, which falls outside of the detector acceptance. 
To ensure consistency between quantities defined in SR and CRs, making the distribution in \ptmiss from $\PW\PZ$ and $\PZ\PZ$ events in the CRs analogous to those in the SR, the additional leptons present in the $\PW\PZ$ CR ($\Pell_3$) and $\PZ\PZ$ CR ($\Pell_3$ and $\Pell_4$) are removed from the calculation of \ptmiss. 
This also increases the statistical precision of these CRs, because \ptmiss increases and more events pass the $\ptmiss > 65\GeV$ requirement.

\begin{figure}[htbp!]
\centering
\includegraphics[width=\cmsFigWidth]{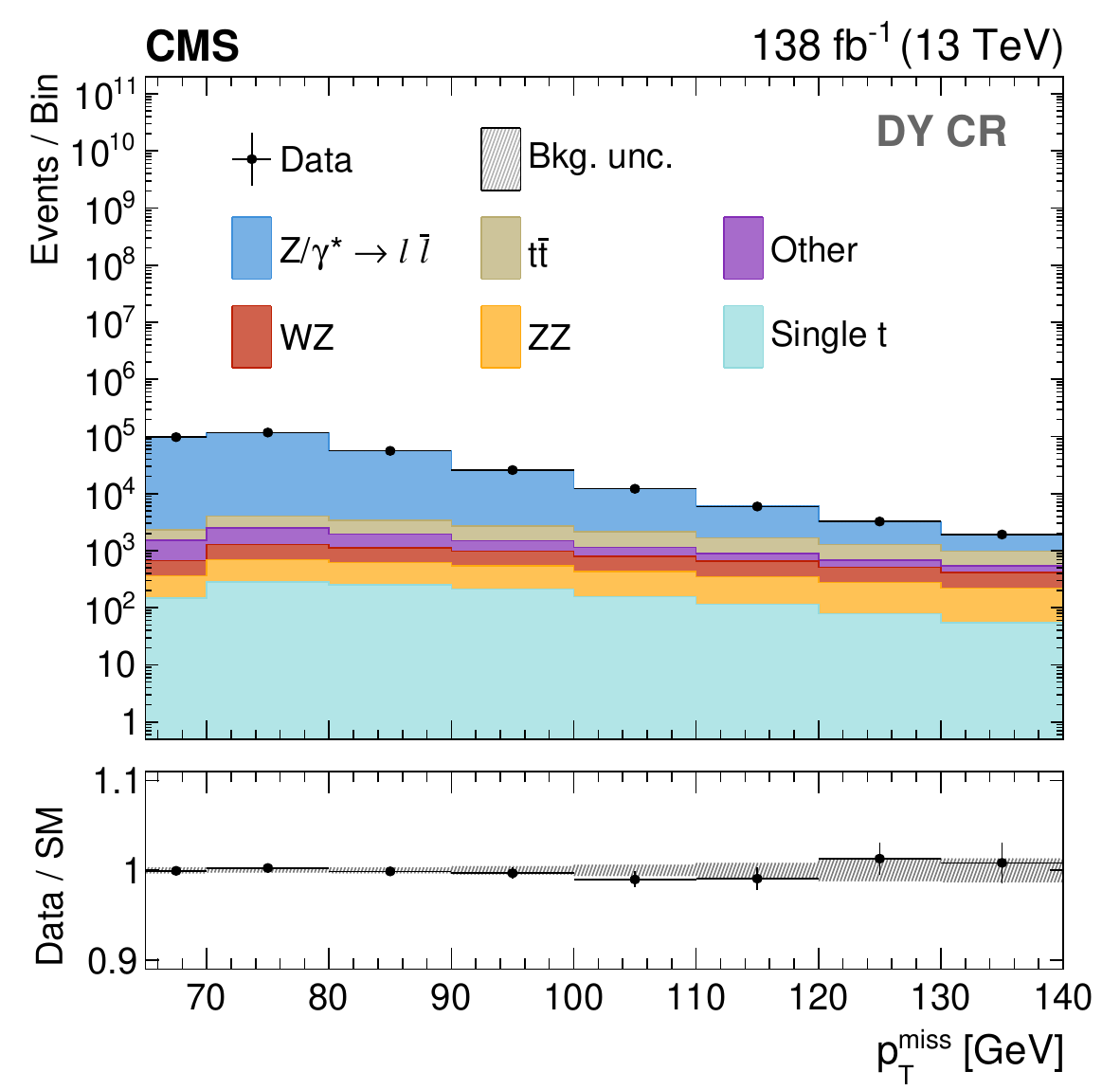}
\includegraphics[width=\cmsFigWidth]{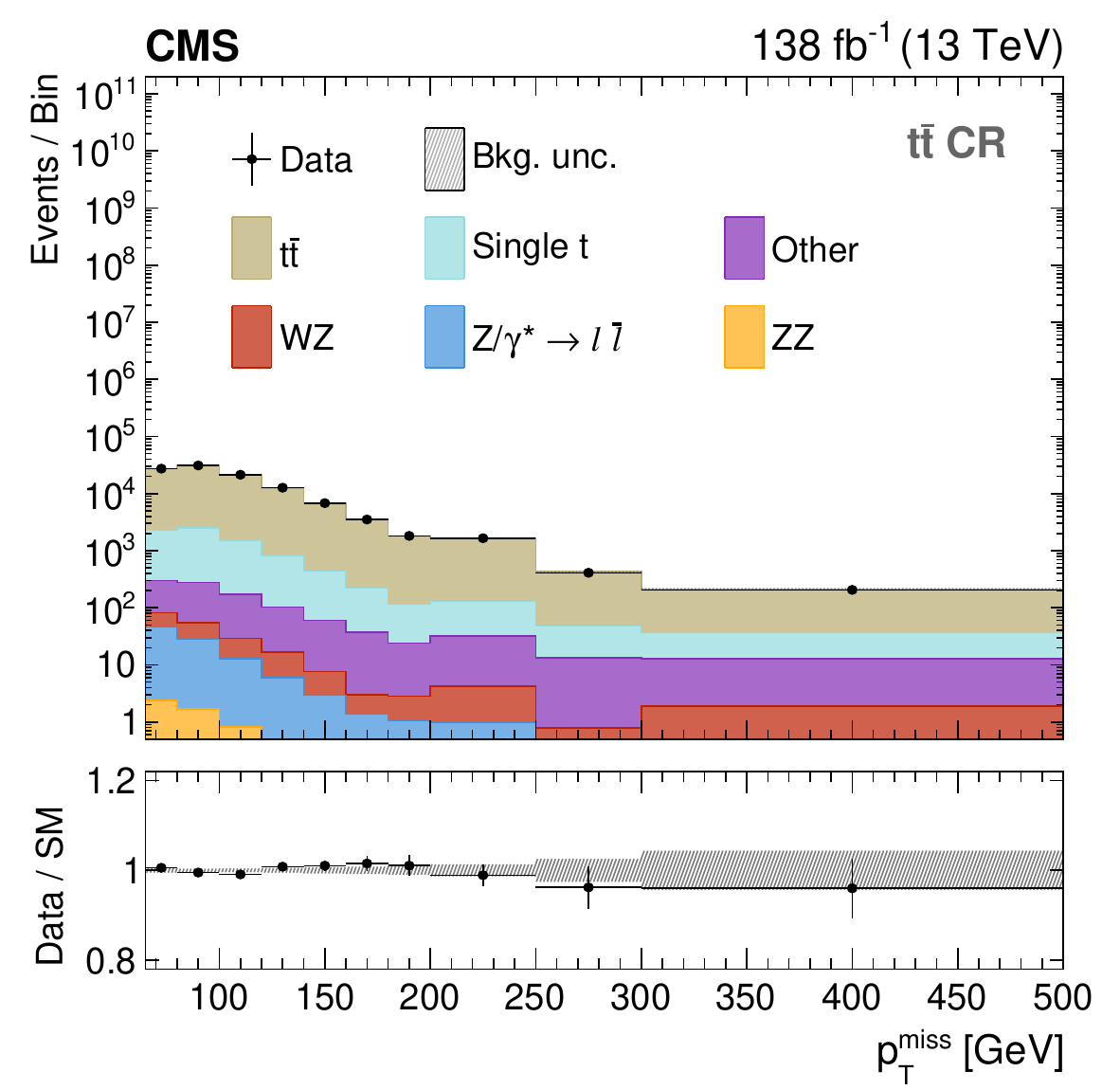}
\includegraphics[width=\cmsFigWidth]{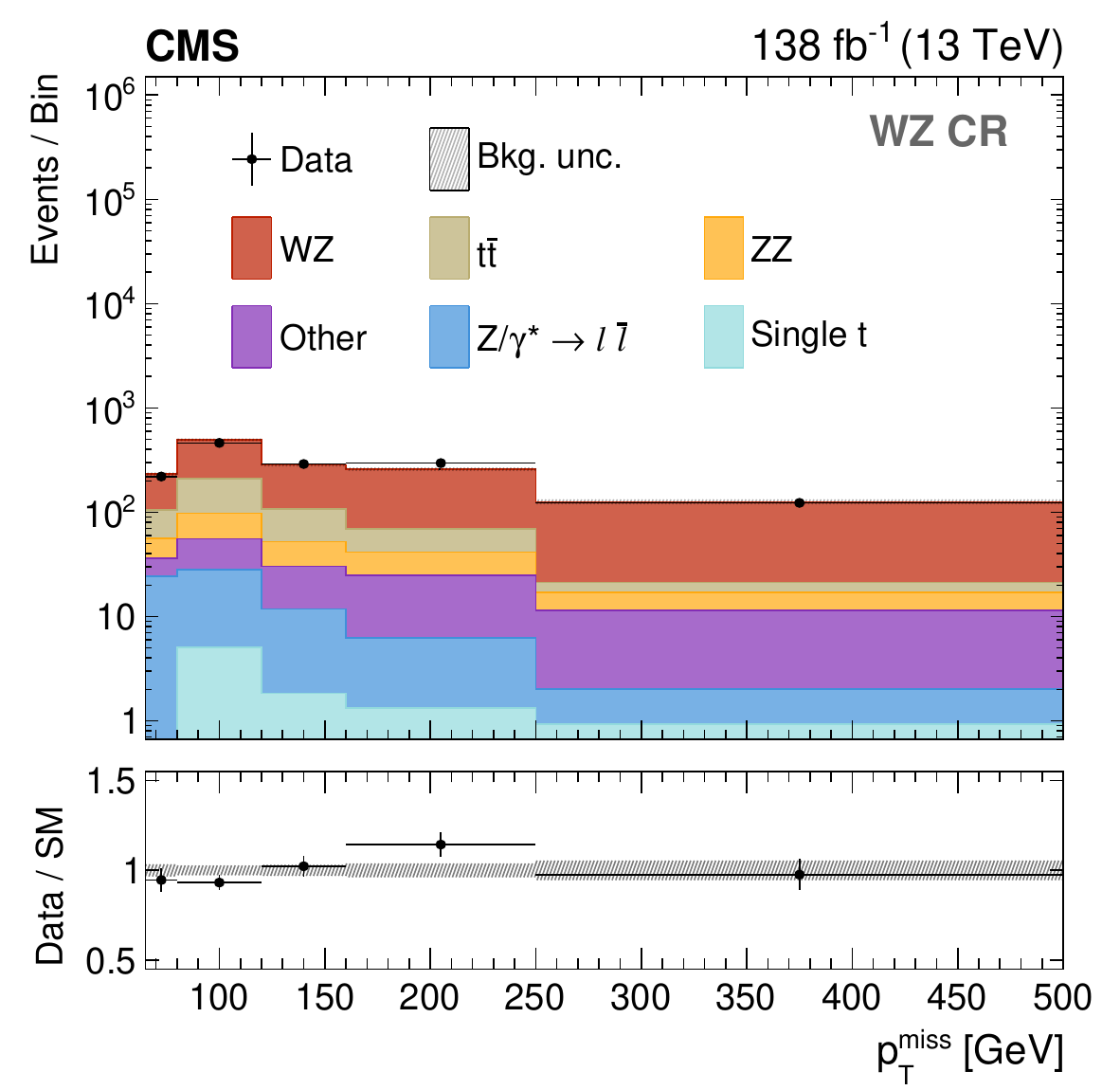}
\includegraphics[width=\cmsFigWidth]{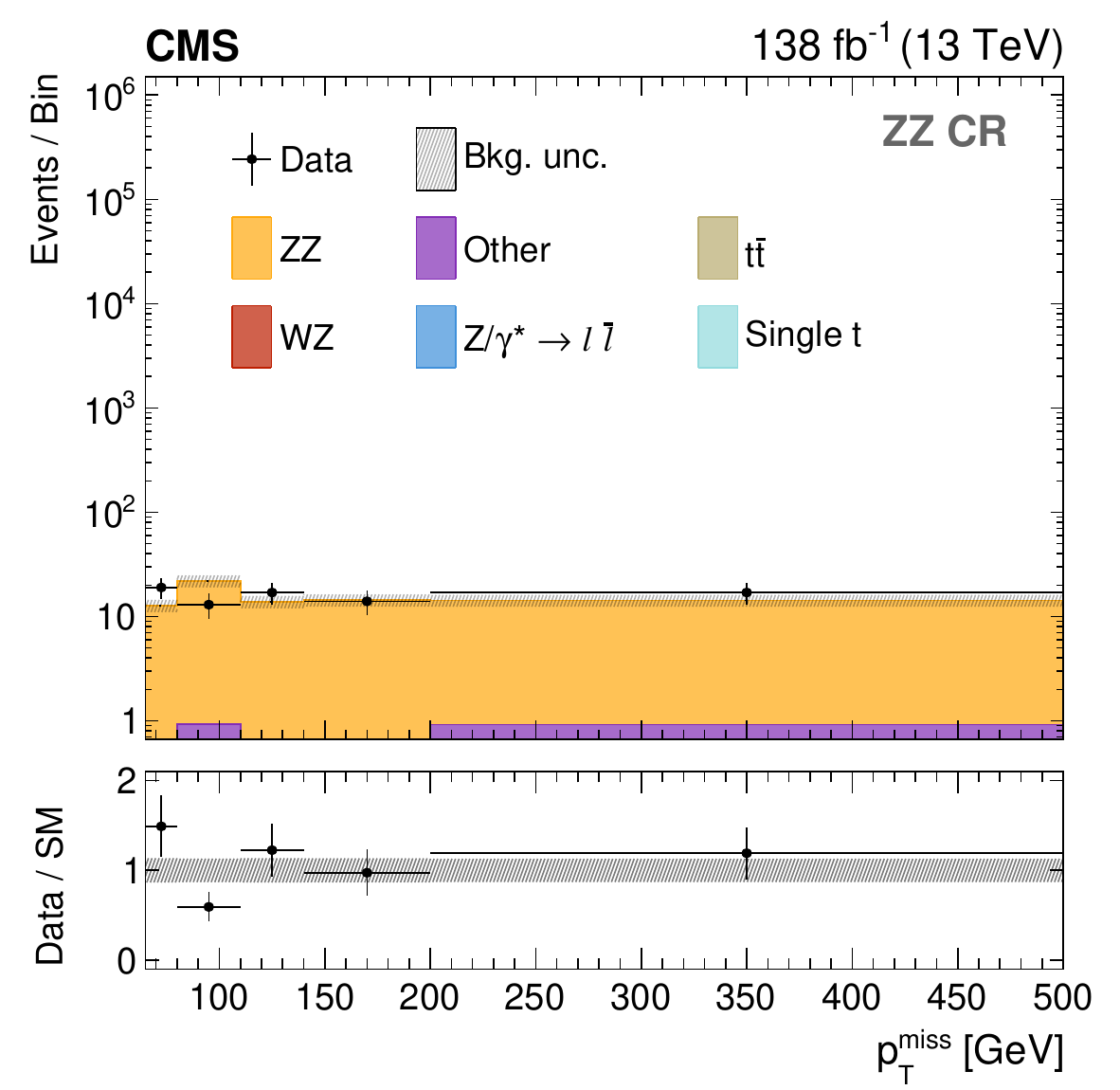}
\caption{Distributions in \ptmiss for the DY (upper left), \ttbar (upper right), $\PW\PZ$ (lower left), and $\PZ\PZ$ (lower right) CRs. In the $\PW\PZ$ and $\PZ\PZ$ CRs, \ptmiss is obtained by removing the additional leptons from the calculation. 
The distributions are shown after performing a background-only fit in the \ptmiss distributions of all CRs. 
The last bin includes the overflow, except for the DY CR where $\ptmiss < 140\GeV$. 
The lower panels show the post-fit values and uncertainties of the ratio between the observed data and the predicted SM backgrounds. The various background processes are represented by filled histograms. The data are shown as black circles, where the vertical bars represent the statistical uncertainty and the 
horizontal bars indicate the bin width.}
\label{Fig:Met-Distribution_CRs}
\end{figure}

A comparison is made in each CR between data and simulation in the distributions of all MLP input variables to verify that the modeling is adequate for the different background processes.
Because of its importance, the distributions in \ptmiss after a background-only fit to the data in the four CRs is shown in Fig.~\ref{Fig:Met-Distribution_CRs}. A maximum likelihood fit is performed on the \ptmiss distribution, combining only the four CRs. The same group of systematic uncertainties described in Section~\ref{sec:systematics} is included in this fit. We observe agreement in shape and normalization between data and simulation within the uncertainties.

\begin{figure}[htbp!]
\centering
\includegraphics[width=\cmsFigWidth]{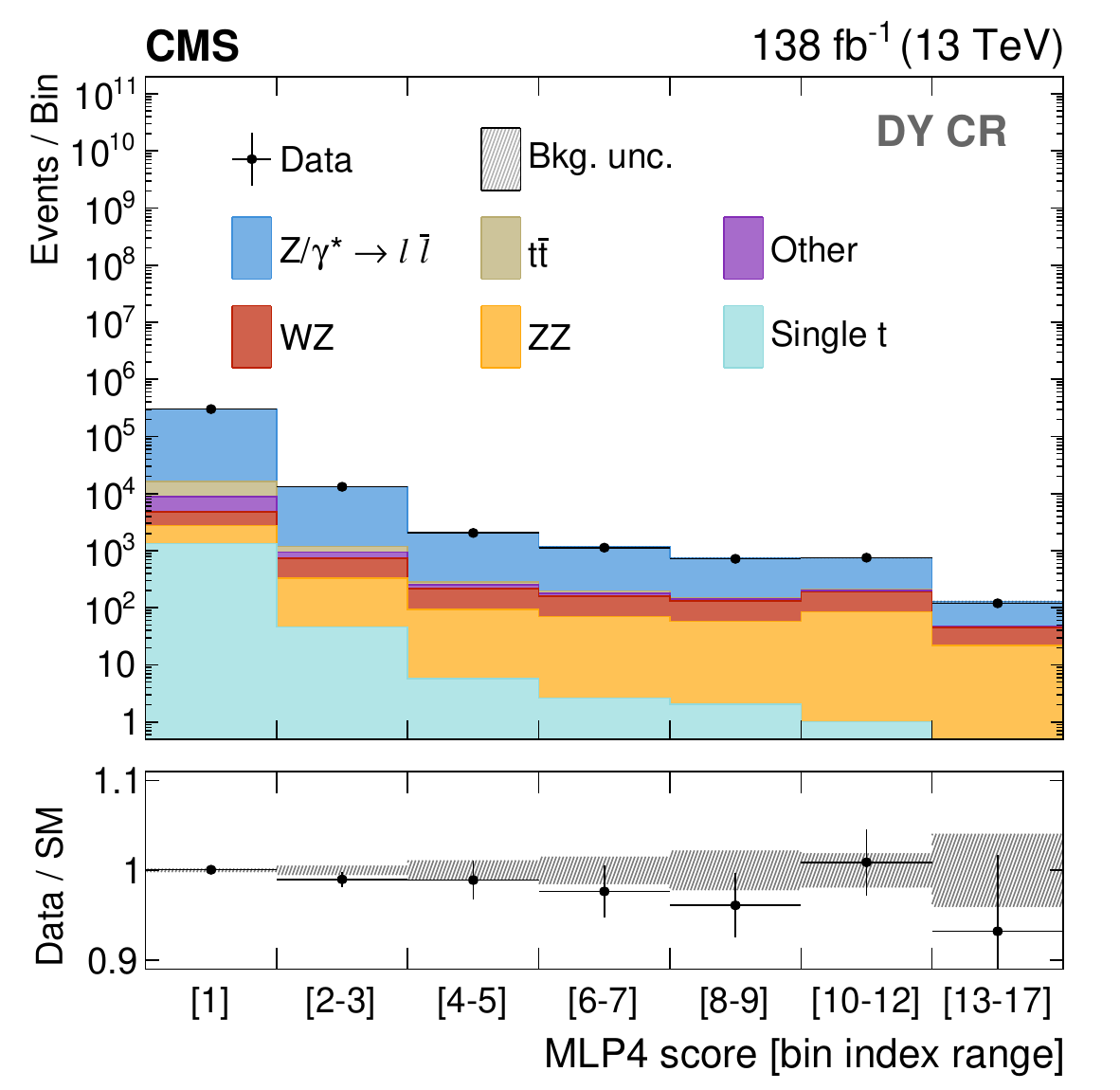}
\includegraphics[width=\cmsFigWidth]{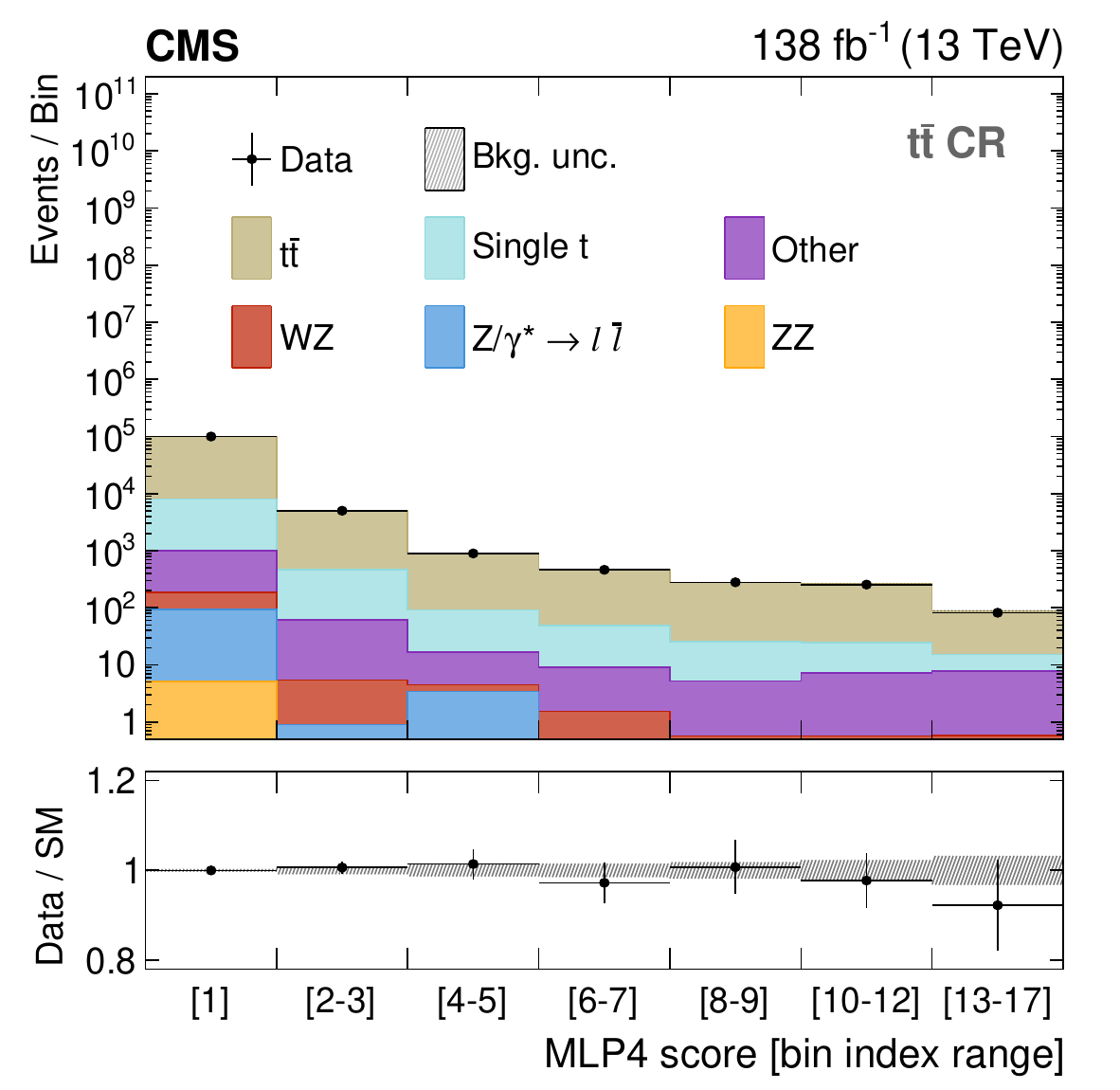}
\includegraphics[width=\cmsFigWidth]{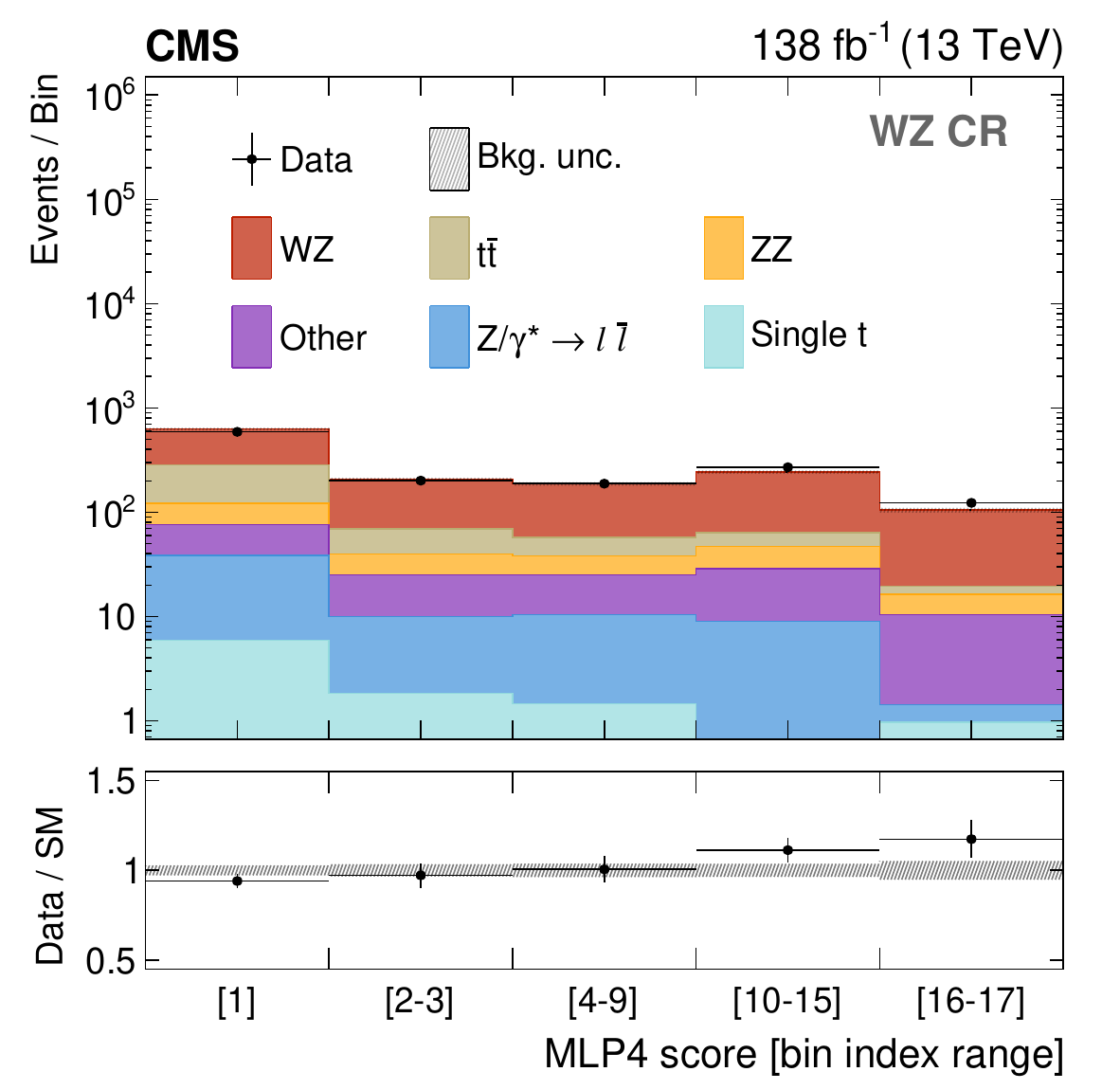}
\includegraphics[width=\cmsFigWidth]{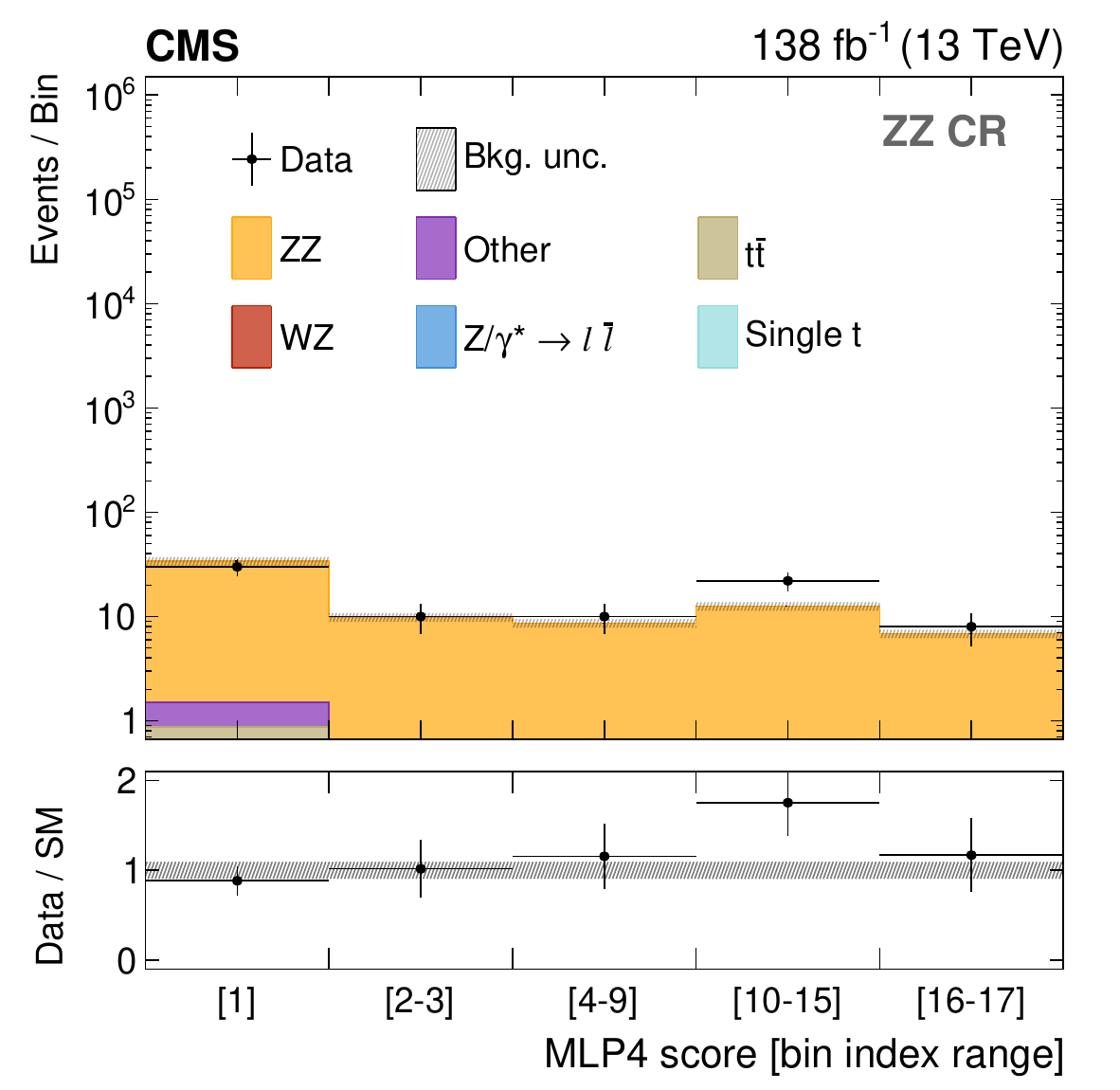}
\caption{Distributions in the MLP4 score for the DY (upper left), \ttbar (upper right), $\PW\PZ$ (lower left), and $\PZ\PZ$ (lower right) CRs. The distributions are shown after performing a background-only fit in the MLP4 score distributions of all CRs. The labels on the horizontal axes indicate the MLP4 intervals that define a given bin. The lower panels show the post-fit values and uncertainties of the ratio between the observed data and the predicted SM backgrounds. The various background processes are represented by filled histograms. The data are shown as black dots, where the vertical bars represent the statistical uncertainty and the 
horizontal bars indicate the bin width.}
\label{Fig:MLP4-Distribution_CRs}
\end{figure}

The distributions in the MLP4 score are compared between data and simulated events in the CRs to verify the modeling of the search variable. For this check, the granularity of the MLP4 score as defined for the SR is too high, because of the smaller statistical precision in the CRs, especially for the $\PW\PZ$ and $\PZ\PZ$ CRs. We merge several of the defined MLP4 bins in order to obtain a reasonable number of observed events in each bin of the MLP4 distributions in the CRs. 
The resulting distributions are shown in Fig.~\ref{Fig:MLP4-Distribution_CRs} after a background-only fit to the data in the CRs, which was executed in a similar way as done for \ptmiss. We observe a good modeling of the MLP4 score by the simulated SM processes that constitute the 
largest backgrounds in this search.

\section{Systematic uncertainties}
\label{sec:systematics}

Several uncertainty sources of both experimental and theoretical nature are taken into consideration when performing the combined maximum likelihood fit described in Section~\ref{sec:results}. These uncertainties can affect some or all simulated processes depending on their source. Most uncertainties in this analysis affect both the shape and normalization of the signal and background processes, except the uncertainty on the integrated luminosity, which only affects the overall rate. Various degrees of correlation for nuisance parameters arising from analogous subsidiary measurements performed in different data-taking conditions are considered.

Among the theoretical uncertainties affecting the shape and the normalization of all relevant processes are the renormalization and factorization scales in the ME calculation. Variations in these scales by a typical factor of two, in either one of them or both of them simultaneously, estimate the uncertainty from missing higher-order corrections, that can lead to changes of the differential cross sections as a function of various kinematic variables. This kinematic shape effect is then transferred to the MLP4 score and is thus considered in the analysis as a general shape effect for different signal and background processes. A similar procedure is used to estimate the uncertainty in the parton shower for the simulation of initial- and final-state radiation, where the scale of the strong coupling is changed by factors of $2$ and $0.5$. We use per-event weights~\cite{Butterworth:2015oua,Accardi:2016ndt} to estimate the uncertainty in the PDFs used to generate the signal and background processes. 

An uncertainty that is specific to the \ttbar background is related to the NNLO QCD (+ NLO EW) correction of the top quark \pt distribution applied to the NLO QCD generated sample, as described in Section~\ref{sec:samples}. The uncertainty in this correction is estimated by taking the difference between the case where this correction is applied and the one where it is not applied. This uncertainty is one of the largest uncertainties in the simulation of the \ttbar background. Analogously, for the correction applied to the $\PW\PZ$ process that employs a similar level of accuracy as for the \ttbar process (see Section~\ref{sec:samples}), a comparable procedure is used. Because the $\PW\PZ$ process contributes more significantly in the bins with the highest signal concentration, the associated uncertainty has a larger impact on the overall sensitivity than the analogous uncertainty in the \ttbar background, and it also dominates over the uncertainties from the DY-related \ptmiss scale and resolution corrections.

Theoretical uncertainties can substantially reduce the analysis sensitivity in the model-specific interpretation, where the limited accuracy of the LO signal simulation introduces a normalization uncertainty of up to 20\%. However, the experimental systematic uncertainties and the statistical uncertainties are the largest ones in this analysis. The statistical size of the CRs directly affects the normalization uncertainties in the corresponding SM processes. These uncertainties can have a large impact, especially on processes classified by the MLP4 score as more signal-like, such as $\PZ\PZ$ and $\PW\PZ$. The proportion of these normalization uncertainties in the main backgrounds, primarily arising from the CR sample sizes, represents the leading factor determining the sensitivity of this search. Also, the limited size of some of the simulated event samples plays an important role in bins with very high values of the MLP4 score, where the background count is low. The statistical uncertainties from the limited size of simulated samples are modeled with the Barlow--Beeston method~\cite{BARLOW1993219}.

An important group of experimental uncertainty sources is related to the identification, reconstruction, and triggering efficiencies of the lepton candidates~\cite{CMS:2018rym,CMS:2020uim,CMS:2018fks}. These uncertainties also include uncertainties in data-to-simulation correction factors that affect the shape of the various processes. The uncertainties in the reconstruction and identification of individual electron candidates range between 1.0 and 2.5\%, and for muons these amount to about 1.0\% in the identification and 0.5\% in the isolation requirements. The total uncertainty associated with the trigger efficiency is approximately 3\%. Due to an early timing offset in the muon and ECAL trigger signals (so-called L1 trigger prefiring) during certain data-taking periods, a correction in simulation and its corresponding systematic uncertainty were included. Depending on the selection region and the kinematic properties of the specific process, this translates into an uncertainty of a few percent in the normalization of the signal and background processes. 

Uncertainties in the reconstruction of jets as well as in \ptmiss are also relevant, most notably the jet energy scale and resolution, and the unclustered energy component of \ptmiss. The change in normalization from the latter uncertainty can be as large as 10\% for some processes such as DY. This process is particularly sensitive to systematic effects in the misreconstruction of \ptmiss, where nonnegligible values arise from the limited precision in the reconstruction of jets and unclustered particles. 

The \PQb-tagging efficiency is corrected in simulation to match the one observed in data through an event-by-event reweighting~\cite{CMS:2017wtu}. The effect of the corresponding uncertainty on the signal normalization is around 2--3\% on average, while for processes with a larger number of \PQb-tagged jets, such as \ttbar production, it can be as large as 6\%. As expected, the uncertainty related to the \PQb-tagging efficiency in the analysis has primarily an impact on the normalization of the simulated processes.

The uncertainty associated with the scale and resolution correction applied to \ptmiss of the DY process has a direct impact on the most important input variables to the MLP classifier, such as \ptmiss itself and $\mT^{\mathrm{\Pell\Pell},\ptmiss}$. 
This is an important experimental uncertainty for signal hypotheses with small $\pt^{\Pell\Pell}$ that tend to have a slightly lower MLP4 score.

The simulated samples are reweighted to reproduce the distribution in the number of pileup interactions observed in data. A corresponding uncertainty is estimated by recalculating these weights for variations in the total inelastic cross section by $\pm5\%$~\cite{CMS:2018mlc}.

The last experimental systematic source is the integrated luminosity for each data-taking period (2016, 2017, and 2018), accounting for an associated uncertainty of 1.2--2.5\%~\cite{CMS:2021xjt,CMS-PAS-LUM-17-004,CMS-PAS-LUM-18-002}, with an overall uncertainty for 2016--2018 data amounting to 1.6\%. This uncertainty affects the rate of the simulated signal and minor background processes estimated entirely from the simulation. 

The maximum magnitude of the various systematic uncertainties affecting the signal and background processes, averaged across the individual analysis categories and processes, is shown in Table~\ref{Tab:Systematics}.

\begin{table}[htbp!]
	\centering
	\topcaption{Displayed are the maximum variations arising from systematic uncertainties that influence the shape and normalization ($\ln$N) of the signal and background processes. Only sources with an impact larger than 1\% are reported.}
		\begin{tabular}{lcc}
			\hline
			{Systematics}                     & {Type} & {Size} \\
			\hline
            Integrated luminosity                          & $\ln$N        & $<$2.5\%        \\
			Pileup reweighting                             & shape$+\ln$N         & 3.5\%         \\
			\PQb tagging efficiency                        & shape$+\ln$N         & $<$5.9\%      \\
			Electron reconstruction and identification     & shape$+\ln$N         & 3.2\%         \\
			Muon identification and isolation              & shape$+\ln$N         & 3.0\%         \\
			Trigger                                        & shape$+\ln$N         & 2.9\%         \\
            L1 prefiring                                   & shape$+\ln$N         & 1.3\%         \\
			Jet energy scale                               & shape$+\ln$N         & 3.1\%         \\
            Jet energy resolution                          & shape$+\ln$N         & 2.5\%         \\
            Jet pileup identification                      & shape$+\ln$N         & 2.1\%         \\
            Unclustered energy \ptmiss                     & shape$+\ln$N         & 3.3\%         \\
            Scale and resolution \ptmiss DY                & shape$+\ln$N         & 4.5\%         \\
            W boson \pt correction WZ                      & shape$+\ln$N         & 4.4\%         \\
			Top quark \pt correction $\ttbar$              & shape$+\ln$N         & 1.9\%         \\
			Renormalization and factorization scales       & shape$+\ln$N         & $<$20\%       \\
			PDF                                            & shape$+\ln$N         & $<$2.9\%      \\
            Parton shower                                  & shape$+\ln$N         & $<$5.7\%      \\
			\hline
		\end{tabular}
	\label{Tab:Systematics}
\end{table}

\section{Results}
\label{sec:results}

A combined maximum likelihood fit~\cite{CMS:2024onh} across all regions is performed using a binned discriminator obtained from the MLP4 score in the SR and single bins in the CRs. The probability density functions of both hypotheses under consideration (background-only and signal-plus-background) are obtained primarily by means of MC simulation, with appropriate corrections determined from subsidiary measurements in the data. These functions are constructed by employing histogram templates. For the SR, the templates are divided into 17 bins according to the MLP4 score output, as indicated in Section~\ref{Subsec:MVA}.

In each of the CRs, a single-bin subsidiary measurement is performed, which creates a link between processes across all the regions. This way, the normalizations of the dominant background processes as obtained from the corresponding CRs are corrected in all regions simultaneously, especially in the SR. The single-bin subsidiary measurements constrain the parameters connected to the systematic uncertainties and correct any bias in the normalization of the process arising due to limited knowledge and modeling. In this way, effects related to either the limited accuracy in the cross sections of the background processes, or small calibration differences between MC samples and the data can be addressed.

\begin{figure}[htbp]
\centering
\includegraphics[width=1.0\textwidth]{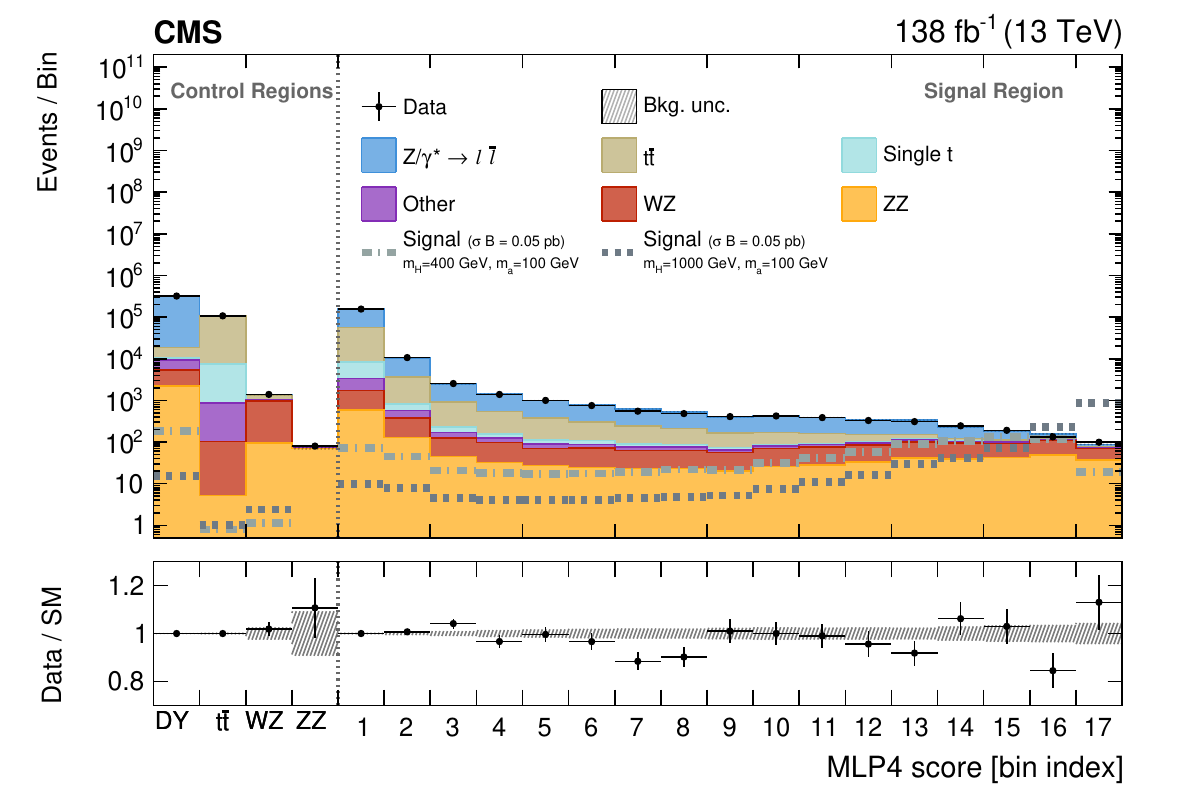}
\caption{Main statistical discriminant of the analysis used to extract the signal after having performed a background-only fit to the observed data. The left side of the upper panel, separated by a vertical dotted line from the right side, shows the four CRs used to estimate the normalization of the main background processes entering the SR. The right side shows the full MLP4 score distribution in the SR used to discriminate between signal and background. The various background processes are represented by filled histograms. The data points are shown as black dots, with vertical bars representing the statistical uncertainty and horizontal bars indicating the bin width, while the signal scenarios under consideration are represented with a dashed-dotted line. The benchmark signal cross section is set to 0.05\unit{pb} for proper visualization purposes. The figure comprises the full combination of all final states and categories for the full data set. The lower panel shows the post-fit values and uncertainties of the ratio between the observed data and the predicted SM background.}
\label{Fig:FinalDiscriminat}
\end{figure}

The above description is depicted in Fig.~\ref{Fig:FinalDiscriminat}, where the different contributions are shown after performing a background-only fit to the data. The left panel (first four bins) shows the CRs of the analysis, while the remaining bins show the search intervals for the MLP4 discriminator in the SR,
with the $y$ axis indicating the total number of events selected. 
During the fit the normalization of each relevant background process is adjusted to match the data in these CRs. With this procedure, the normalizations of the largest background processes are changed, increasing by 10 and 8\% for DY and $\ttbar$, respectively. The relative uncertainties on the estimated normalizations for DY and $\ttbar$ are 0.5\% and 0.8\%, respectively.
The changes in normalization obtained for $\PW\PZ$ and $\PZ\PZ$ are much larger, resulting in an increase of 101 and 113\%, respectively, of the total yields. The relative uncertainties on the estimated normalizations for $\PW\PZ$ and $\PZ\PZ$ are 6\% and 10\%, respectively. The larger normalization factors for $\PW\PZ$ and $\PZ\PZ$ are understood and associated with the specific phase space probed, in particular, the simultaneous requirements of a large \ptmiss and the presence of b jets. Previous CMS analyses, such as Ref.~\cite{CMS:2020ulv}, have already observed discrepancies in the modeling of $\PW\PZ$ and $\PZ\PZ$ processes in phase-space regions with similar \ptmiss requirements. Given that this search further requires a minimum number of b jets, and in view of the well-known limitations in the description of diboson production with additional b quarks, it is well motivated to expect an additional, non-negligible contribution to the overall mismodeling from this source. Note additionally that, in this work, the diboson cross sections are taken primarily at NLO accuracy, since the fit can naturally absorb residual corrections to the overall normalization of background processes independently of the initial pre-fit values. Several higher-order QCD calculations~\cite{Cascioli:2014yka,Grazzini:2017ckn,Grazzini:2018owa} have nevertheless demonstrated that corrections beyond this order can be sizable for the $\PW\PZ$ and $\PZ\PZ$ processes. 

The results of the background-only fit reveal a good level of agreement between the SM prediction and the observed data. The maximum local significance for an excess of events over the SM prediction is below 0.7 standard deviation and occurs for the signal hypothesis with $m_{\PH}=2000$\GeV and $m_{\Pa}=400$\GeV. Although the dominant backgrounds are DY and \ttbar production in the dilepton final state, the $\PW\PZ$ and $\PZ\PZ$ processes tend to be classified as being more signal-like. For these two background processes, events featuring a reconstructed $\PZ$ boson accompanied by substantial \ptmiss are more prevalent, for the reasons discussed in Section~\ref{sec:samples}. This leads to the uncertainties in the normalization of these EW processes playing an essential role, becoming the most important systematic uncertainties in this analysis. As seen in the signal distributions, the MLP4 score tends to have higher values for signal events. This is illustrated for two signal mass configurations with $m_{\PH}=400$ and 1000\GeV for $m_{\Pa}=100$\GeV. Depending on the mass difference $m_{\PH}-m_{\Pa}$, the concentration of signal events at high values of the MLP4 discriminator will be higher or lower for large and small values of the mass difference, respectively.

\subsection{Upper limits on the signal cross sections}
\label{subsec:xsec-limits}

The results from the maximum likelihood fits are used to set upper limits at 95\% confidence level (\CL) on the product of the cross section and branching fraction 
$\sigma\BR = \sigma (\pp \to \bbbar\PH) \allowbreak \BR(\PH \to \PZ \Pa ) \BR(\PZ \to \Pl\PAl) \BR(\Pa \to \PGc\PAGc)$. The exclusion limits are calculated using the asymptotic approximation of the \CLs method~\cite{ALRead_2002,Cowan:2010js}. 

Under the assumption that the narrow-width approximation is valid for all resonances involved in the decay chain, the computed upper limits only depend on the masses of the new resonances involved, i.e., $m_{\PH}$ and $m_{\Pa}$. Those limits can be translated into constraints on the parameters of a concrete model via the dependence of the cross section and branching fractions.

\begin{figure}[t]
\centering
\includegraphics[width=\textwidth]{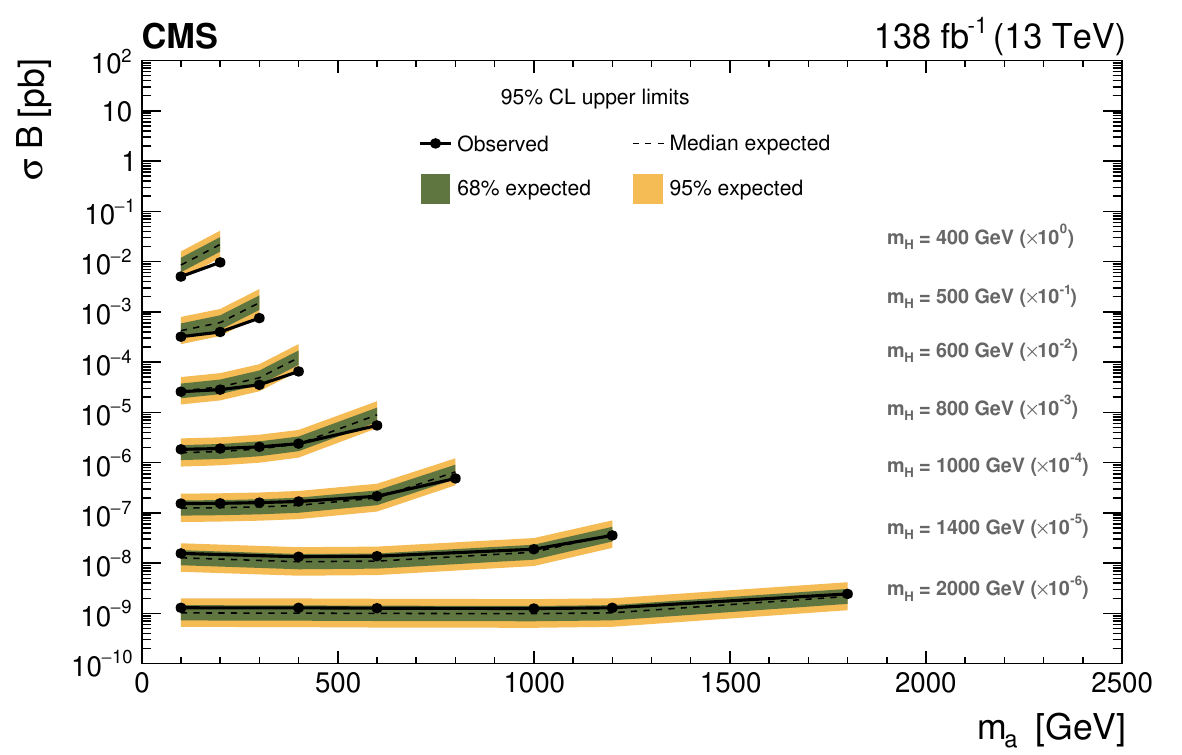}
\caption{ Observed and expected upper limits at 95\% \CL on the product of the signal cross section and branching fractions $\sigma\BR$. The dependence of the limits on the pair $(m_{\PH},m_{\Pa})$ has been accommodated into various 1D projections for a fixed value of $m_{\PH}$, where the corresponding limits have been scaled by an arbitrary factor ($\times 10^{-n}$) for visualization purposes. The $y$ axis contains the obtained cross section upper limit for the various combinations, whereas the $x$ axis exhibits the dependence on the mass of the pseudoscalar. The solid and dashed lines correspond to the observed and median expected limits, respectively, while the green and yellow bands indicate the regions that contain 68\% and 95\% of the expected upper limits. 
}
\label{Fig:MILimits1D}
\end{figure}

Figure~\ref{Fig:MILimits1D} shows the observed and expected upper limits on the product of the cross section and branching fraction as function of the masses of the two resonances. The two-dimensional (2D) dependence has been unrolled into a one-dimensional graphic for better illustration, i.e. each limit band depicted along the $y$ axis corresponds to a fixed value of \mH, while the dependency on \ma is given along the $x$ axis. 
The expected and observed limits at 95\% \CL range from $10^{-3}$ to $10^{-2}$\unit{pb}.
The discrepancies with a statistical deviation greater than one standard deviation, observed in the limits for the signal hypothesis with lower \mH and $\mH - \ma$, are associated with the small deficit seen around bin 16 in the MLP4 distribution in SR, since these signal hypotheses have their peak in bins just below bin 17.

The analysis is sensitive to semi-boosted scenarios, namely when the mass difference $m_{\PH}-m_{\Pa}$ is around 1\TeV. For very small mass differences, the selection requirements imposed on leptons and \ptmiss tend to have an important impact on the signal acceptance, as it usually happens in compressed scenarios~\cite{ATLAS:2018lkz,CMS:2021edw}, where the energy of the decay products is not sufficient to pass the minimal \pt requirements on the physics objects. 
On the other hand, when the mass difference is too large, the leptonic products arising from the decay of the \PZ bosons start to overlap in the $(\eta,\phi)$ space, thus making the successful reconstruction and identification of the leptonic pair increasingly less likely.

\subsection{Interpretation in the 2HDM+\texorpdfstring{\Pa}{a} context}
\label{subsec:2hdma-interpretation}

The results on $\sigma\BR$ are translated into constraints on the parameters of the 2HDM+$\Pa$ model using the benchmark scenario specified in Eq.~\eqref{Eq:BenchmarkChoice}. We verified that the effect of the parameter variations is negligible in the kinematic distributions of the signal samples in the probed parameter space. A summary of the latest experimental constraints on the 2HDM+$\Pa$ model, obtained from searches targeting a variety of final states, can be found in Refs.~\cite{CMS:2024zqs,ATLAS:2023rvb}.

Four relevant parameters are simultaneously varied to estimate the dependence of $\sigma\BR$: $m_{\PH}$, $m_{\Pa}$, $\tan\beta$, and $ \sin\theta$. The calculations are performed using MG5 and the \textsc{MadWidth}~\cite{Alwall:2014bza} functionality within it. Several 2D projections of the constraints are constructed by varying two of the above parameters while keeping the rest fixed to their values in Eq.~\eqref{Eq:BenchmarkChoice}. A 2D interpolation procedure is used to obtain the values for mass configurations that are not generated and parameter values for which the cross sections have not been calculated. In cases where one of the masses $m_{\PH}$ or $m_{\Pa}$ is fixed, the values $m_{\PH}=800$\GeV and $m_{\Pa}=300$\GeV are chosen. For each of these configurations in the 2HDM+a, we delimit regions favored by DM relic density calculations, obtained by assuming a velocity-averaged annihilation cross section to be in the range $\langle \sigma v \rangle =$ (2--4)$\times 10^{-26}\,\mathrm{cm^{3}/s}$. Those calculations were done using the \textsc{MadDM} tool~\cite{Backovic:2013dpa}. The above interval is constructed by varying the canonical value obtained for the thermally averaged cross section at the time of freeze-out, resulting from the observed DM relic abundance~\cite{Planck:2015fie}. This value corresponds to $\langle \sigma v \rangle \approx 3 \times 10^{-26}\,\mathrm{cm^{3}/s}$~\cite{Steigman:2012nb}. We then choose to vary it by one third up and down to construct a conservative coverage interval around the central value, similarly as suggested in Ref.~\cite{Tunney:2017yfp}. 

\begin{figure}[htbp]
\centering
\includegraphics[width=0.49\textwidth]{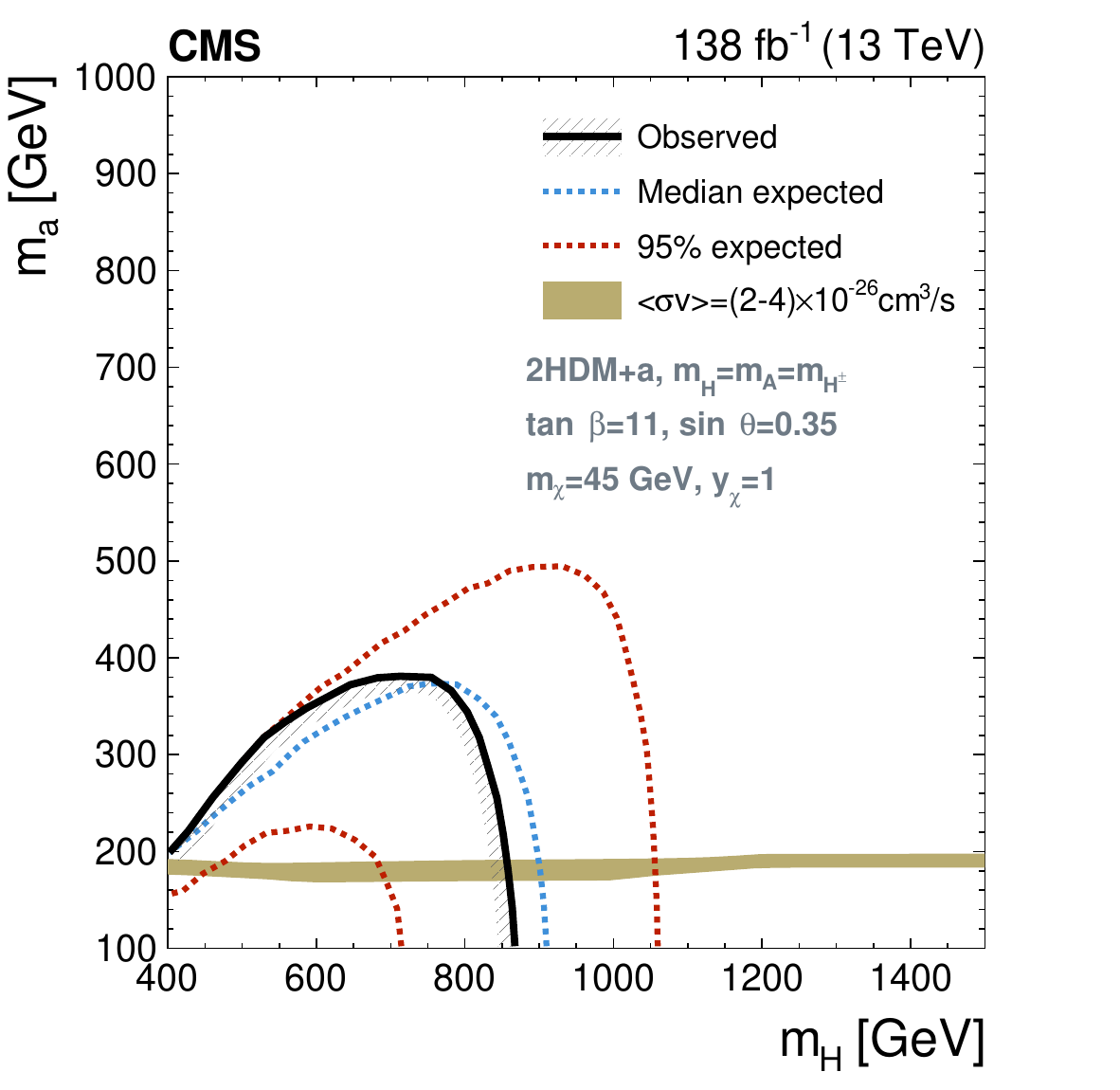}
\includegraphics[width=0.49\textwidth]{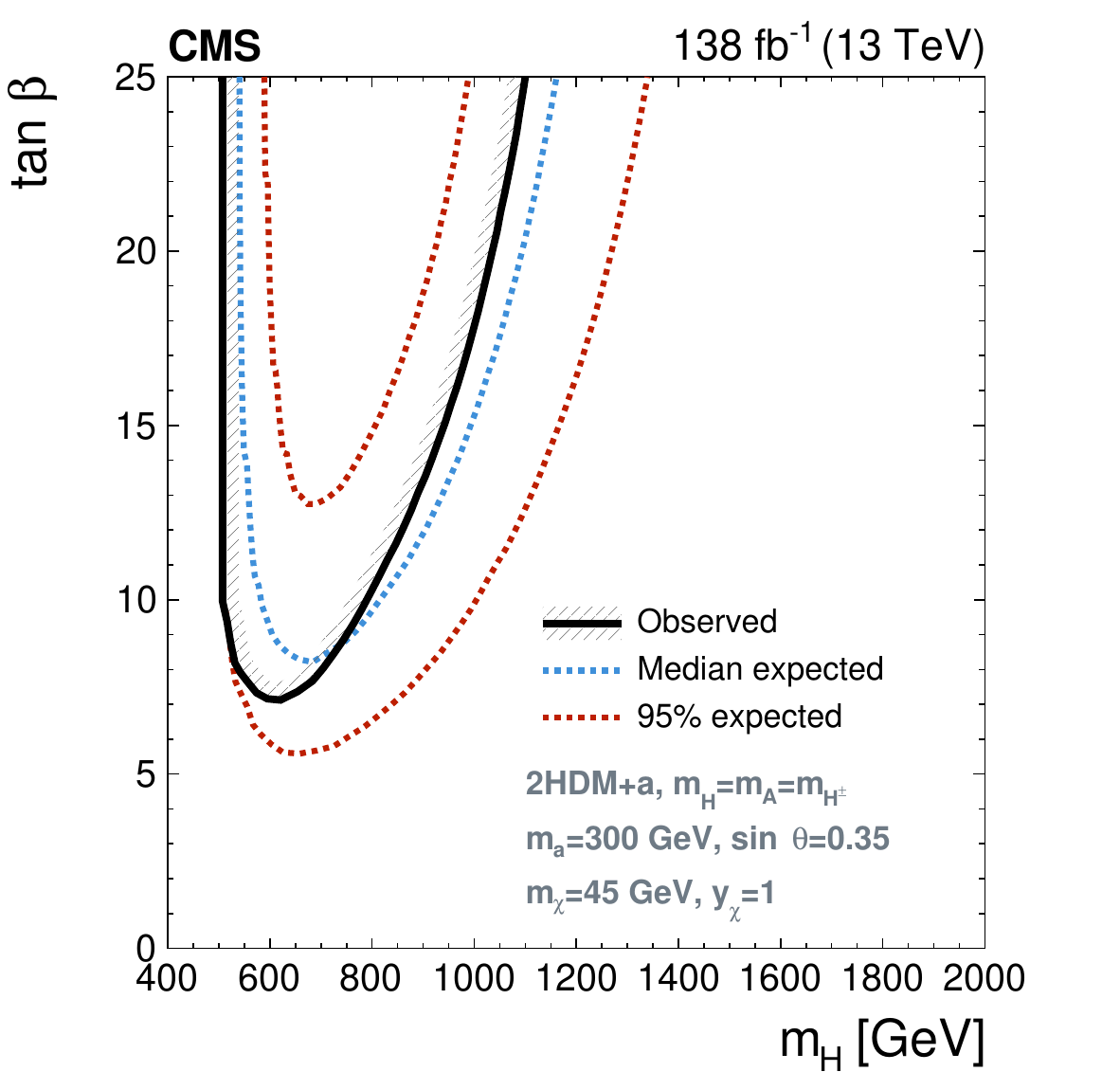}
\includegraphics[width=0.49\textwidth]{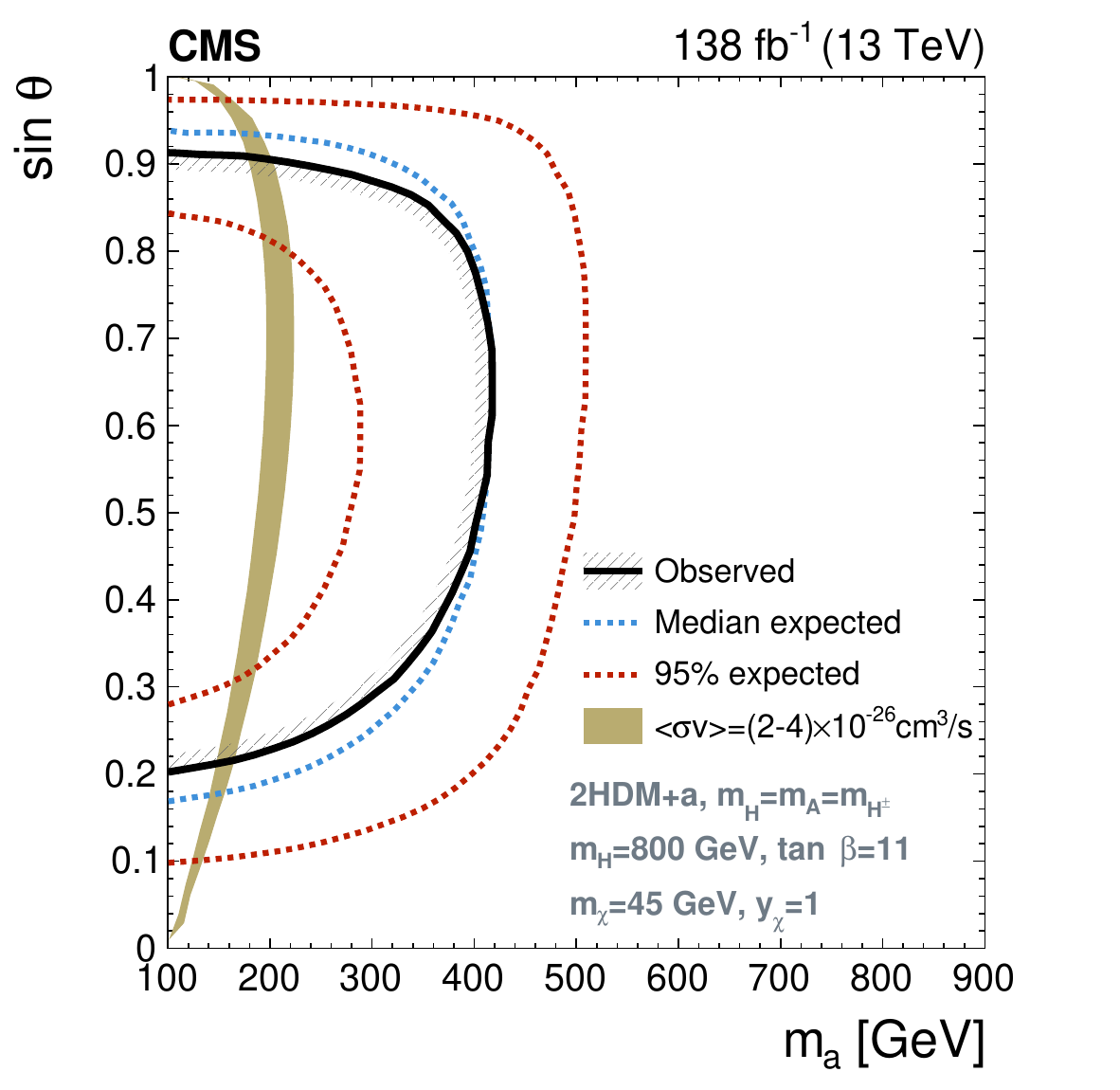}
\includegraphics[width=0.49\textwidth]{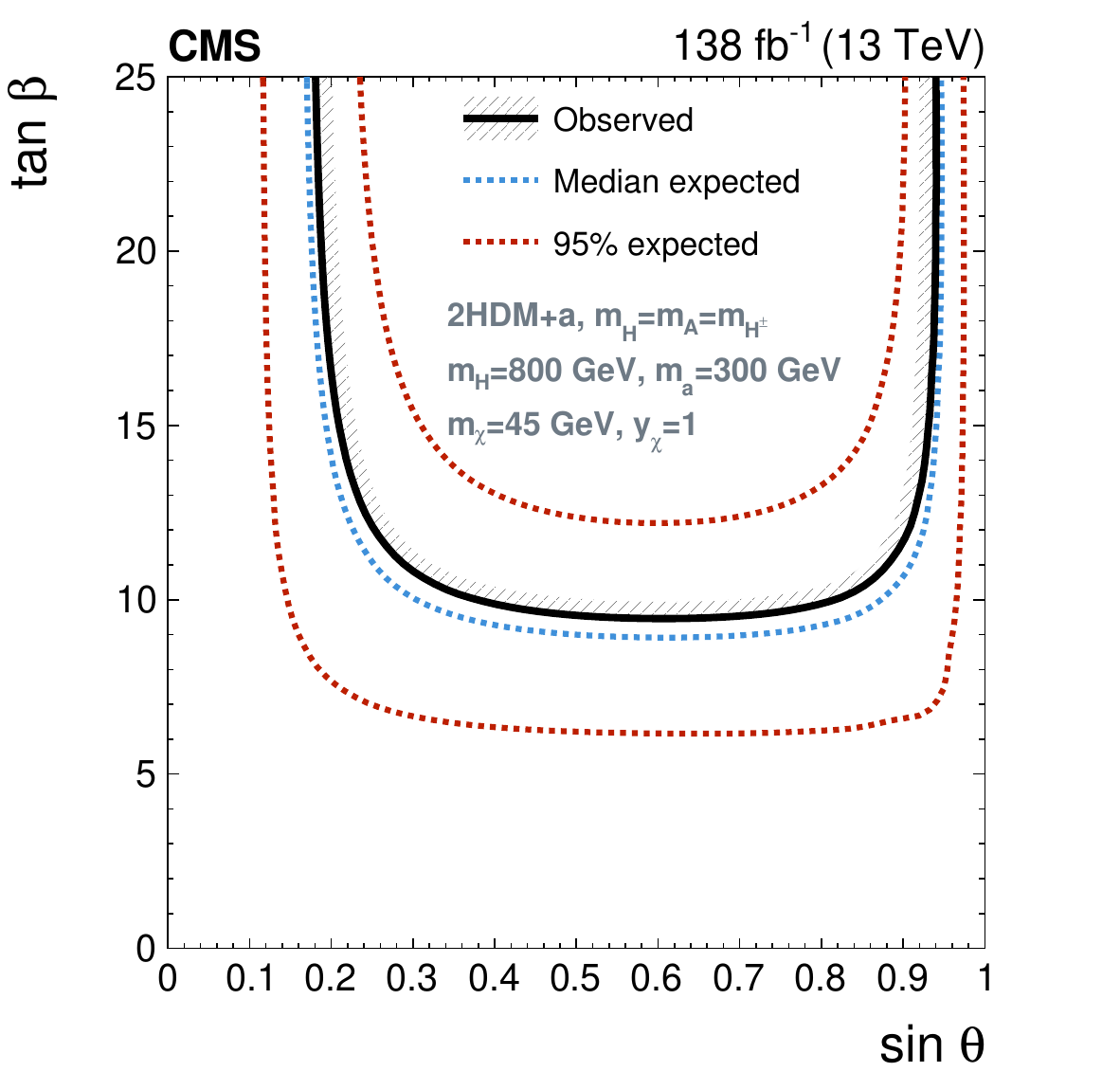}
\caption{Excluded regions in the parameter space of the 2HDM+a. The solid lines encompass the observed excluded regions, the dashed blue lines the expected, and the red-dotted lines indicate the regions that contain 95\% of the expected exclusion limits. Projections are presented for the ($m_{\PH}$, $m_{\Pa}$) plane (upper left), ($m_{\PH}$, $\tan\beta$) plane (upper right), ($m_{\Pa}$, $\sin\theta$) plane (lower left), and ($\tan\beta$, $\sin\theta$) plane (lower right), for fixed values of the parameters in Eq. \eqref{Eq:BenchmarkChoice}. The olive green band represents the allowed region as estimated from $\langle \sigma v \rangle =$ (2--4)$\times 10^{-26}\,\mathrm{cm^{3}/s}$, which covers a range around the central value required by the observed DM relic. The cases where this curve is not visible in the figures correspond to the scenario where the preferable values of $\tan\beta$ for this range of  $\langle \sigma v \rangle$ fall beyond the threshold ($\tan\beta>25$) depicted in the projections.}
\label{Fig:Exclusions-2HDMa}
\end{figure}

In Fig.~\ref{Fig:Exclusions-2HDMa}, four projections of the exclusion regions in the various 2D planes are shown. In the ($m_{\PH}$, $m_{\Pa}$) projection (upper left plot), one can observe that the analysis is able to exclude masses of the heavy scalar of up to 900\GeV for small masses of the pseudoscalar mediator. This is consistent with what was observed in a previous CMS search~\cite{CMS:2020ulv}. The previous study was not optimized for the associated production of $\PH$ boson with b quarks, and thus could not access scenarios with high values of \tanb. 
In the ($m_{\PH}$, $\tan\beta$) projection (upper right plot), it is visible that the analysis is mainly sensitive to regions of the parameter space of large values of $\tan\beta$. For the chosen parameter values in this article, the analysis is able to exclude masses of the heavy scalar close to 1.1\TeV for values of $\tan\beta \sim 25$. In this particular projection, one can observe that, given that $m_{\Pa}$ is fixed to 300\GeV, values of $m_{\PH}$ below 500\GeV are almost inaccessible for this analysis. This is caused by both the dramatic reduction of phase space, thus impacting $\mathcal{B}(\PH \to \PZ\Pa )$, and lower sensitivity to less boosted mass configurations. 
The same behavior regarding very compressed mass configurations can be seen in the ($m_{\Pa}$, $\sin\theta$) projection (lower left plot), where one observes that the analysis is not capable of excluding masses of the pseudoscalar larger than 400\GeV, given that the heavy scalar mass has been fixed to 800\GeV. However, the analysis is able to cover a large part of the parameter space for $m_{\Pa} < 400$\GeV, because of a multiplicative factor proportional to $\sin^{2}\theta \cos^{2}\theta$ in the partial decay width of $\PH \to \PZ a$. 
This projection shows the capability of this analysis to exclude regions of the parameter space that are favored by the cosmological estimations of the DM relic density, assuming the 2HDM+$\Pa$ and considering the preference towards high values of $\tan\beta$~\cite{Tunney:2017yfp}. This analysis excludes the low-$m_{\Pa}$ regions for a very broad range of values of $\sin\theta$, leaving only corners with very small or very large values of $\sin\theta$ uncovered. 
The last projection (lower right plot) onto the ($\tan\beta$, $\sin\theta$) plane shows the sensitivity of this analysis to intermediate values of $\sin\theta$, where a large part of the parameter space is excluded for $\tan\beta>9$ for fixed values of $m_\Pa$ and $m_\PH$. As $\sin\theta$ approaches 0 or 1, the resulting limits deteriorate significantly, reflecting the corresponding suppression of the $\PH \to \PZ\Pa$ and $\Pa\to\PGc\PAGc$ decay modes. A change in the values of $m_\Pa$ and $m_\PH$, most importantly in the mass difference, causes an increase or decrease of the excluded region for larger or smaller mass differences, respectively.

\section{Summary}
\label{sec:summary}

The first dedicated search for dark matter (DM) with the CMS experiment has been presented, where the DM particles are produced through the production of a heavy neutral Higgs boson (\PH) in association with a bottom quark-antiquark (\bbbar) pair, followed by the decay $\PH\to\PZ\Pa$ with $\Pa\to\PGc\PAGc$, where \Pa is a pseudoscalar mediator and $\PGc\PAGc$ denote the DM particle and antiparticle. A data set of proton-proton collisions at a center-of-mass energy of 13\TeV, corresponding to an integrated luminosity of 138\fbinv, is analyzed. 

This analysis exploits for the first time a signature involving a \PZ boson decaying into a pair of electrons or muons combined with requirements on the number of \PQb jets and the amount of missing transverse momentum. A discriminator obtained with machine-learning techniques is used to separate the signal from background events. The multivariate classifier is trained to reach a high level of discrimination across a broad range of kinematic variations that arise from the different configurations in which the \PZ boson and the DM mediator are produced.

No signs of DM production via the channel investigated here have been observed. The results are presented in terms of limits on the product of signal cross section and branching fractions for the decays $\PH\to\PZ\Pa$, $\Pa\to\PGc\PAGc$, and $\PZ\to\Pl\PAl$, where \Pl denotes a charged lepton. The 95\% confidence level upper limits for the production cross section  branching fraction of the new particles vary between $10^{-2}$ and $10^{-3}$\unit{pb} for heavy Higgs masses between 400 and 2000\GeV, respectively. Constraints on the parameter space of a two Higgs doublet model plus a pseudoscalar (2HDM+\Pa) benchmark are derived. Exclusion regions in two-dimensional planes formed from four relevant 2HDM+\Pa parameters are shown. The results are compared with expectations for this model in the context of cosmological predictions, in particular with the constraints arising from the thermally averaged cross section at the time of freeze-out, which are dictated by the observed DM relic abundance. The experimental results exclude a significant region of the parameter space preferred by those predictions for some relevant scenarios of the 2HDM+\Pa model.

\begin{acknowledgments}
We congratulate our colleagues in the CERN accelerator departments for the excellent performance of the LHC and thank the technical and administrative staffs at CERN and at other CMS institutes for their contributions to the success of the CMS effort. In addition, we gratefully acknowledge the computing centers and personnel of the Worldwide LHC Computing Grid and other centers for delivering so effectively the computing infrastructure essential to our analyses. Finally, we acknowledge the enduring support for the construction and operation of the LHC, the CMS detector, and the supporting computing infrastructure provided by the following funding agencies: SC (Armenia), BMBWF and FWF (Austria); FNRS and FWO (Belgium); CNPq, CAPES, FAPERJ, FAPERGS, and FAPESP (Brazil); MES and BNSF (Bulgaria); CERN; CAS, MoST, and NSFC (China); MINCIENCIAS (Colombia); MSES and CSF (Croatia); RIF (Cyprus); SENESCYT (Ecuador); ERC PRG, TARISTU24-TK10 and MoER TK202 (Estonia); Academy of Finland, MEC, and HIP (Finland); CEA and CNRS/IN2P3 (France); SRNSF (Georgia); BMFTR, DFG, and HGF (Germany); GSRI (Greece); NKFIH (Hungary); DAE and DST (India); IPM (Iran); SFI (Ireland); INFN (Italy); MSIT and NRF (Republic of Korea); MES (Latvia); LMTLT (Lithuania); MOE and UM (Malaysia); BUAP, CINVESTAV, CONACYT, LNS, SEP, and UASLP-FAI (Mexico); MOS (Montenegro); MBIE (New Zealand); PAEC (Pakistan); MES, NSC, and NAWA (Poland); FCT (Portugal); MESTD (Serbia); MICIU/AEI and PCTI (Spain); MOSTR (Sri Lanka); Swiss Funding Agencies (Switzerland); MST (Taipei); MHESI (Thailand); TUBITAK and TENMAK (T\"{u}rkiye); NASU (Ukraine); STFC (United Kingdom); DOE and NSF (USA).

\hyphenation{Rachada-pisek} Individuals have received support from the Marie-Curie program and the European Research Council and Horizon 2020 Grant, contract Nos.\ 675440, 724704, 752730, 758316, 765710, 824093, 101115353, 101002207, 101001205, and COST Action CA16108 (European Union); the Leventis Foundation; the Alfred P.\ Sloan Foundation; the Alexander von Humboldt Foundation; the Science Committee, project no. 22rl-037 (Armenia); the Fonds pour la Formation \`a la Recherche dans l'Industrie et dans l'Agriculture (FRIA) and Fonds voor Wetenschappelijk Onderzoek contract No. 1228724N (Belgium); the Beijing Municipal Science \& Technology Commission, No. Z191100007219010, the Fundamental Research Funds for the Central Universities, the Ministry of Science and Technology of China under Grant No. 2023YFA1605804, the Natural Science Foundation of China under Grant No. 12061141002, 12535004, and USTC Research Funds of the Double First-Class Initiative No.\ YD2030002017 (China); the Ministry of Education, Youth and Sports (MEYS) of the Czech Republic; the Shota Rustaveli National Science Foundation, grant FR-22-985 (Georgia); the Deutsche Forschungsgemeinschaft (DFG), among others, under Germany's Excellence Strategy -- EXC 2121 ``Quantum Universe" -- 390833306, and under project number 400140256 - GRK2497; the Hellenic Foundation for Research and Innovation (HFRI), Project Number 2288 (Greece); the Hungarian Academy of Sciences, the New National Excellence Program - \'UNKP, the NKFIH research grants K 131991, K 133046, K 138136, K 143460, K 143477, K 146913, K 146914, K 147048, 2020-2.2.1-ED-2021-00181, TKP2021-NKTA-64, and 2021-4.1.2-NEMZ\_KI-2024-00036 (Hungary); the Council of Science and Industrial Research, India; ICSC -- National Research Center for High Performance Computing, Big Data and Quantum Computing, FAIR -- Future Artificial Intelligence Research, and CUP I53D23001070006 (Mission 4 Component 1), funded by the NextGenerationEU program (Italy); the Latvian Council of Science; the Ministry of Education and Science, project no. 2022/WK/14, and the National Science Center, contracts Opus 2021/41/B/ST2/01369, 2021/43/B/ST2/01552, 2023/49/B/ST2/03273, and the NAWA contract BPN/PPO/2021/1/00011 (Poland); the Funda\c{c}\~ao para a Ci\^encia e a Tecnologia, grant CEECIND/01334/2018 (Portugal); the National Priorities Research Program by Qatar National Research Fund; MICIU/AEI/10.13039/501100011033, ERDF/EU, "European Union NextGenerationEU/PRTR", and Programa Severo Ochoa del Principado de Asturias (Spain); the Chulalongkorn Academic into Its 2nd Century Project Advancement Project, the National Science, Research and Innovation Fund program IND\_FF\_68\_369\_2300\_097, and the Program Management Unit for Human Resources \& Institutional Development, Research and Innovation, grant B39G680009 (Thailand); the Kavli Foundation; the Nvidia Corporation; the SuperMicro Corporation; the Welch Foundation, contract C-1845; and the Weston Havens Foundation (USA).
\end{acknowledgments}

\bibliography{auto_generated}\cleardoublepage \appendix\section{The CMS Collaboration \label{app:collab}}\begin{sloppypar}\hyphenpenalty=5000\widowpenalty=500\clubpenalty=5000\cmsinstitute{Yerevan Physics Institute, Yerevan, Armenia}
{\tolerance=6000
V.~Chekhovsky, A.~Hayrapetyan, V.~Makarenko\cmsorcid{0000-0002-8406-8605}, A.~Tumasyan\cmsAuthorMark{1}\cmsorcid{0009-0000-0684-6742}
\par}
\cmsinstitute{Institut f\"{u}r Hochenergiephysik, Vienna, Austria}
{\tolerance=6000
W.~Adam\cmsorcid{0000-0001-9099-4341}, J.W.~Andrejkovic, L.~Benato\cmsorcid{0000-0001-5135-7489}, T.~Bergauer\cmsorcid{0000-0002-5786-0293}, K.~Damanakis\cmsorcid{0000-0001-5389-2872}, M.~Dragicevic\cmsorcid{0000-0003-1967-6783}, C.~Giordano, P.S.~Hussain\cmsorcid{0000-0002-4825-5278}, M.~Jeitler\cmsAuthorMark{2}\cmsorcid{0000-0002-5141-9560}, N.~Krammer\cmsorcid{0000-0002-0548-0985}, A.~Li\cmsorcid{0000-0002-4547-116X}, D.~Liko\cmsorcid{0000-0002-3380-473X}, I.~Mikulec\cmsorcid{0000-0003-0385-2746}, J.~Schieck\cmsAuthorMark{2}\cmsorcid{0000-0002-1058-8093}, R.~Sch\"{o}fbeck\cmsAuthorMark{2}\cmsorcid{0000-0002-2332-8784}, D.~Schwarz\cmsorcid{0000-0002-3821-7331}, M.~Sonawane\cmsorcid{0000-0003-0510-7010}, W.~Waltenberger\cmsorcid{0000-0002-6215-7228}, C.-E.~Wulz\cmsAuthorMark{2}\cmsorcid{0000-0001-9226-5812}
% ...
\par}
\vskip\cmsinstskip
\dag:~Deceased\\
$^{1}$Also at Yerevan State University, Yerevan, Armenia\\
$^{2}$Also at TU Wien, Vienna, Austria\\
% ...
\end{sloppypar}
\end{document}